%% file: main.tex
\documentclass[aps,prl,twocolumn,superscriptaddress,10pt]{revtex4-2}
\usepackage{graphicx}
\usepackage{hyperref}
\usepackage{natbib}
\usepackage{xfrac}
\usepackage{xcolor}
\usepackage[english]{babel}
\usepackage[]{lineno}

\newcommand{\hu}{\affiliation{Department of Physics, Humboldt-Universität zu Berlin, 12489 Berlin, Germany}}
\newcommand{\fbh}{\affiliation{Ferdinand-Braun-Institut (FBH), 12489 Berlin, Germany}}

\begin{document}

\author{Alok Gokhale}
\hu
\author{Marco E. Stucki}
\hu 
\fbh
\author{Malo Bezard}
\hu 
\fbh
\author{Maarten H. van der Hoeven}
\hu
\author{Kilian Unterguggenberger}
\hu 
\author{Gregor Pieplow}
\hu
\author{Tommaso Pregnolato}
\hu 
\fbh
\author{Julian M. Bopp}
\altaffiliation[Present address: ]{Fraunhofer-Institut für Nachrichtentechnik, Heinrich-Hertz-Institut, 10587 Berlin, Germany}
\hu 
\fbh
\author{Laura Orphal-Kobin}
\hu
\author{Tim Schröder}
\email[Corresponding author: ]{tim.schroeder@physik.hu-berlin.de}
\hu
\fbh

\title[Cavity Paper]{Frequency multiplexing of Purcell-enhanced optically coherent NV centers}

\begin{abstract}
    Integrating optically coherent quantum emitters into nanophotonic structures is key to scalable quantum networks. Realizing them with nitrogen-vacancy (NV) centers in diamond is hindered by the degradation of optical coherence accompanying nanofabrication. Here, we overcome this challenge by integrating multiple NV centers into a monolithic diamond photonic crystal nanocavity while preserving near-lifetime-limited optical coherence. We observe linewidths as narrow as 19 MHz, enabling the frequency multiplexing of spectrally distinct NV centers within a common cavity mode, while maintaining charge-state stability over several minutes under resonant excitation. Tuning the cavity into resonance with the NV zero-phonon line enhances the optical emission, consistent with Purcell enhancement predicted by finite-element simulations. Additionally, individual cavities can be independently brought into resonance with NV zero-phonon lines on the same chip. These results combine bulk-like optical coherence with frequency multiplexing and cavity-enhanced emission, establishing a scalable nanophotonic platform for NV-based quantum network nodes.
\end{abstract}

\maketitle

\input{main_text}

\section{Acknowledgements}
The authors gratefully acknowledge Nir Rotenberg, Jacob Ewaniuk and Tristan Austin for fruitful discussions. The funding for this project was provided by the European Research Council (ERC, Starting Grant project no. 851810 - QUREP), the German Federal Ministry of Research, Technology and Space (BMFTR, projects QR.X [KIS6QK4001], QR.N [16KIS2185], QPIS [16KISQ032K], QPIC-1 [13N15858], Dinoquant [13N14921]), the Einstein Foundation Berlin (Einstein Research Unit on Quantum Devices), and the Innovation Fund Denmark (project EQUAL [4356-00007B]).

\section{Author contributions}
A.G.: Investigation (lead), Formal analysis (lead), Methodology (equal), Visualization (lead), Project administration (support), Writing – original draft (lead).
M.S.: Investigation (supporting), Methodology (equal).
M.B.: Formal analysis (supporting), Methodology (equal).
K.U.: Investigation (supporting), Formal analysis (supporting), Methodology (equal).
G.P.: Formal analysis (supporting).
M.H.: Investigation (supporting).
T.P.: Investigation (supporting), Methodology (equal).
J.B.: Software (lead), Methodology (equal) Project administration (supporting).
L.O.: Conceptualization (supporting), Methodology (equal), Project administration (supporting), Supervision (supporting).
T.S.: Conceptualization (lead), Funding acquisition (lead), Resources (lead), Supervision (lead), Project administration (lead).
All authors: Writing - review \& editing (equal).

\section{Data Availability}
The data that support the findings of this study are available from the corresponding author upon reasonable request.

\section{Competing Interests}
The authors declare no competing interests

\section{AI Usage Declaration}
The authors acknowledge the use of LLMs (ChatGPT by OpenAI) during manuscript preparation to assist with language editing, manuscript structure, scientific wording and LaTeX formatting. The tool was also used for generation of Python code for analyzing and plotting data. All outputs were critically reviewed and verified by the authors, who take full responsibility for the scientific content and conclusions. Generative AI was not used to generate experimental data or perform experiments.

\bibliography{main_ref}

\input{methods}

\end{document}


\author{Alok Gokhale}
\hu
\author{Marco E. Stucki}
\hu 
\fbh
\author{Malo Bezard}
\hu 
\fbh
\author{Maarten H. van der Hoeven}
\hu
\author{Kilian Unterguggenberger}
\hu 
\author{Gregor Pieplow}
\hu
\author{Tommaso Pregnolato}
\hu 
\fbh
\author{Julian M. Bopp}
\altaffiliation[Present address: ]{Fraunhofer-Institut für Nachrichtentechnik, Heinrich-Hertz-Institut, 10587 Berlin, Germany}
\hu 
\fbh
\author{Laura Orphal-Kobin}
\hu
\author{Tim Schröder}
\email[Corresponding author: ]{tim.schroeder@physik.hu-berlin.de}
\hu
\fbh

\title{\texorpdfstring
{Supplementary notes for \\[1cm] Frequency multiplexing of Purcell-enhanced optically coherent NV centers}
{Supplementary notes for Frequency multiplexing of Purcell-enhanced optically coherent NV centers}
}

\maketitle

\input{supplement_text}
\bibliography{supplement_ref}

%% file: main_text.tex
\begin{figure*}[!ht]
    \centering
    \includegraphics[]{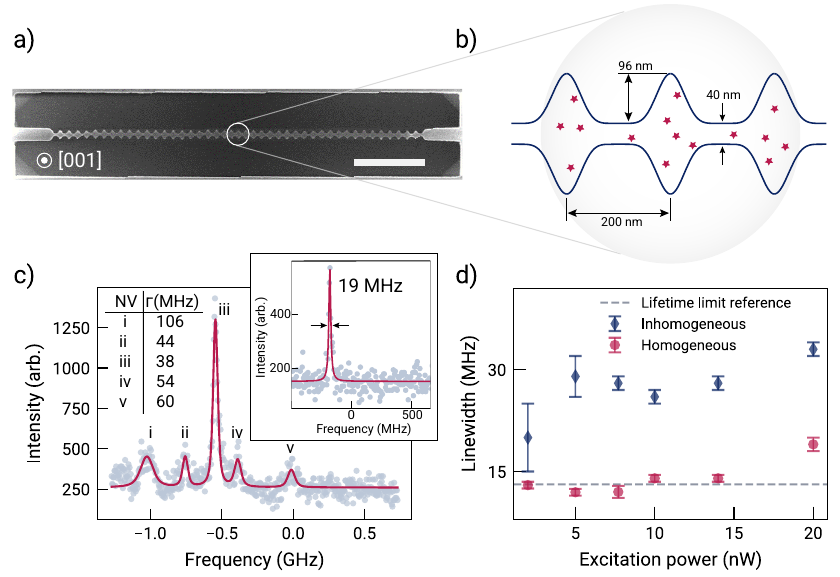}
    \caption{Frequency-multiplexing of optically coherent NV centers in a Sawfish nanocavity. a) Scanning electron micrograph of a representative Sawfish nanocavity. The white circle highlights the region addressed by the diffraction-limited excitation spot. The scale bar is 2\,µm. The long-axis of the cavity is parallel to the [100] crystal direction. b) Enlarged schematic of the region highlighted in a). The shaded circle has a diameter of 500\,nm. Red stars show an example random spatial distribution of 15 NV centers within the illuminated region, based on the expected NV density. c) Inhomogeneous linewidths of cavity-embedded NVs measured when the cavity is off-resonant with the NV ZPL. The main plot shows the integrated signal from approximately 200 PLE scans at an excitation power of 40\,nW over 2.5 \,min. Several NV resonances are visible with their full-width at half-maximum (FWHM), $\Gamma$, given in the inset table. The inset plot shows the integrated PLE spectrum at a excitation power of  2\,nW, recorded over more then 6\,min. A linewidth of 19\,MHz is determined using a Voigt fit. The frequency axes are given relative to 470.541\,THz. d) Power-dependent linewidth measurements. Blue diamonds indicate the inhomogeneous linewidths, each measured over 6\,min. The error bars indicate the standard error of the fitted linewidth. Red circles represent homogeneous linewidth values extracted using a Monte-Carlo approach, with error bars representing the \(\text{3}\sigma\) confidence interval. The grey dashed line indicates the lifetime-limited linewidth.}
    \label{fig:fig1}
\end{figure*}

\begin{figure*}[ht]
    \centering
    \includegraphics[width=\linewidth]{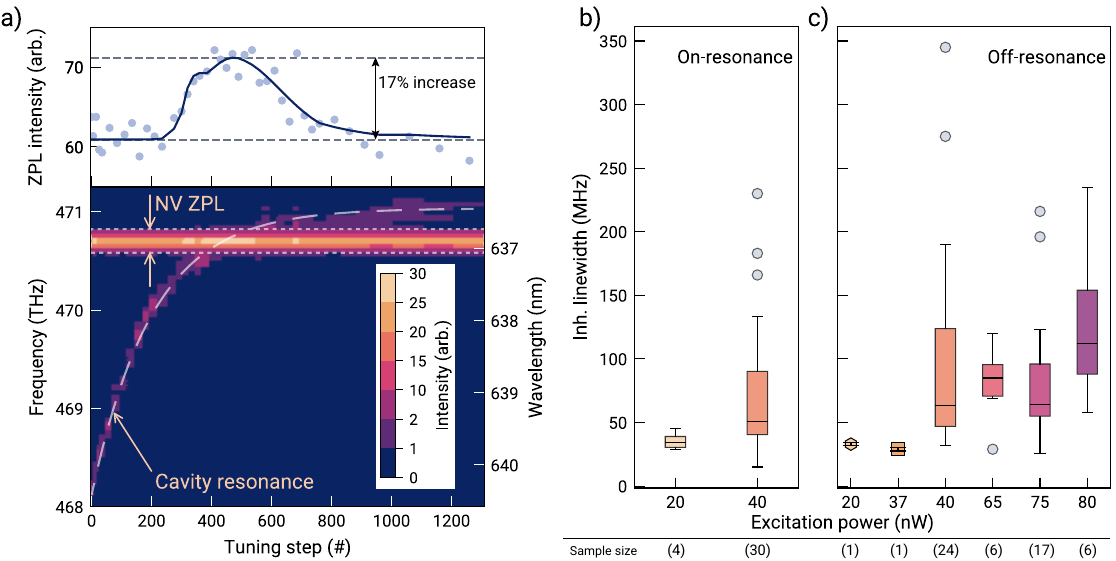}
    \caption{Cavity-enhanced emission and inhomogeneous linewidths of NV centers. a) Bottom: cumulative plot of multiple PL spectra recorded during cavity tuning. The dashed line indicates the cavity-resonance trajectory and the dotted box marks the spectral range of the NV ZPL, limited by the spectrometer resolution. Top: PL intensity of the NV ZPL as the cavity resonance is tuned across the transition. Light blue circles show the measured data. The solid dark blue line is a Voigt fit to the ZPL intensity as a function of the inferred cavity-resonance frequency (see Fig.\,\ref{fig:fig3}a) mapped onto the tuning coordinate shown here. The dashed grey lines represent the fitted baseline and maximum intensity. b-c) Inhomogeneous linewidths with the cavity tuned on resonance (b) and off resonance (c) with the NV centers. Measurements for multiple individual NVs at different excitation powers are fitted with Voigt profiles and shown as box plots. The black lines show the medians, boxes the inter-quartile ranges (IQR) and whiskers extend to 1.5×IQR. Points beyond this range are shown as outliers (grey circles). The yellow hexagon at 20\,nW and the yellow-orange square at 37\,nW in c) indicate individual measurements, with error bars representing  the standard error of the fit.}
    \label{fig:fig2}
\end{figure*}

As quantum technologies transition from proof-of-principle demonstrations toward early generations of practical applications \cite{oh_perspective_2024,aquina_critical_2025,seskir_landscape_2022}, scalable quantum network nodes are becoming increasingly important. Such nodes must store, process, and distribute quantum information while interfacing efficiently with photonic channels. Integrating multiple spectrally distinguishable quantum memories within a single node enables frequency multiplexing \cite{bakr_multiplexed_2025}, increasing communication rates and providing a route towards distributed quantum computing \cite{atature_material_2018,wehner_quantum_2018,nickerson_freely_2014}, secure communication \cite{basso_basset_quantum_2021,ekert_ultimate_2014}, and entanglement-assisted sensing \cite{stas_entanglement-assisted_2026}.

Among the candidate solid-state platforms for realizing scalable quantum network nodes, the negatively charged nitrogen-vacancy center in diamond (NV) combines optically accessible spin qubits with some of the longest coherence times reported in the solid state, ranging from seconds-scale electron-spin coherence to minute-long nuclear-spin memories \cite{balasubramanian_ultralong_2009,maurer_room-temperature_2012,abobeih_one-second_2018,bradley_ten-qubit_2019,yamamoto_tensecond_2026}. Harnessing these properties in an integrated quantum-network architecture requires efficient optical interfacing of NV centers with nanophotonic cavities without compromising their optical coherence. This coherence is essential for generating photon-mediated remote spin-photon entanglement \cite{pompili_realization_2021} and, through sufficiently narrow optical linewidths, enables multiple spectrally distinct NV centers to couple to a common cavity mode. Single NV centers have been coupled to a range of nanocavity platforms \cite{hausmann_coupling_2013,mouradian_rectangular_2017,li_coherent_2015}, with Purcell factors reaching up to 70 \cite{faraon_coupling_2012}. Yet the associated degradation of optical coherence has so far limited the prospects for the integration of these devices into quantum-networking architectures. Bridging this gap between cavity-enhanced light–matter interaction and maintenance of coherence demonstrated in NV-based quantum networks \cite{pompili_realization_2021,stolk_metropolitan-scale_2024} is therefore a key step toward scalable, cavity-integrated network nodes.

Here we demonstrate the integration of multiple spectrally resolvable and optically coherent NV centers into a Sawfish photonic crystal cavity \cite{bopp_sawfish_2024} while preserving near-lifetime-limited optical coherence. We investigate their optical and charge-state properties under resonant excitation. By combining cavity tuning with fluorescence measurements and photoluminescence excitation (PLE) spectroscopy, we characterize optical linewidths, charge-state stability, and emitter–cavity coupling within one device. Finite-element method (FEM) simulations show that the observed fluorescence enhancement is consistent with Purcell-enhanced emission. Finally, our cavity-tuning approach enables the independent alignment of multiple cavities with the NV zero-phonon line (ZPL) transitions on the same chip, complementing the frequency multiplexing of multiple NV centers within a common cavity mode and providing a parallel route towards scalable quantum network nodes based on cavity-integrated NV centers. 

\section{Lifetime-limited NVs in a Sawfish cavity}
To extend lifetime-limited NV optical coherence from diamond nanopillars to nanophotonic cavities, we adapt the methods developed in our earlier work, specifically the choice of a nitrogen-enriched diamond substrate, nanofabrication recipes, and optical control schemes \cite{orphal-kobin_optically_2023}. The cavity design is based on the recently demonstrated Sawfish spin-photon interface \cite{bopp_sawfish_2024, pregnolato_fabrication_2024} and employs a periodic modulation of the waveguide width to vary the effective refractive index of the guided mode, thereby resulting in optical confinement (Fig.\,\ref{fig:fig1}a). The NV centers embedded in this cavity are addressed confocally. Based on the known substitutional nitrogen concentration of about 1\,ppm in our sample \cite{E6_website} and PLE measurements, we estimate that 15(4) NV centers can be addressed optically within the cavity volume, marked by the white circle in Fig.\,\ref{fig:fig1}a, representing the diffraction-limited spot of the confocal setup (see Supplementary Materials). Fig.\,\ref{fig:fig1}b shows one possible spatial configuration of the set of optically addressed NVs. For substrates with a similar nitrogen density as ours, electron-spin coherence times exceeding 100\,µs have been demonstrated \cite{bauch_degoherence_2020}.

At approximately 5\,K, the 15 NV centers in the confocal spot are distributed over a spectral range of about 10\,GHz (see Supplementary materials). The resulting average spectral spacing of about 670\,MHz is much larger than the $\sim$13\,MHz natural linewidth of NVs \cite{orphalkobin_coherent_2025}, enabling individual spectral addressing and frequency multiplexing of the emitters within the cavity mode. Figure \ref{fig:fig1}c shows the inhomogeneous NV linewidths extracted by fitting Voigt profiles to an integrated PLE spectrum recorded at an excitation power of 40\,nW. Four of the five emitters exhibit linewidths within five times the natural linewidth and are stable over integration times exceeding one minute. To examine the coherence attainable at lower excitation powers, we perform additional PLE measurements at 2\,nW (Fig.\,\ref{fig:fig1}c, inset). Notably, these measurements yield an inhomogeneous linewidth as low as 19\,MHz sustained over integration times of several minutes. To our knowledge, this represents the first demonstration of near lifetime-limited inhomogeneous linewidth for a single NV embedded in a nanostructure.
\begin{figure*}[ht]
    \centering
    \includegraphics[]{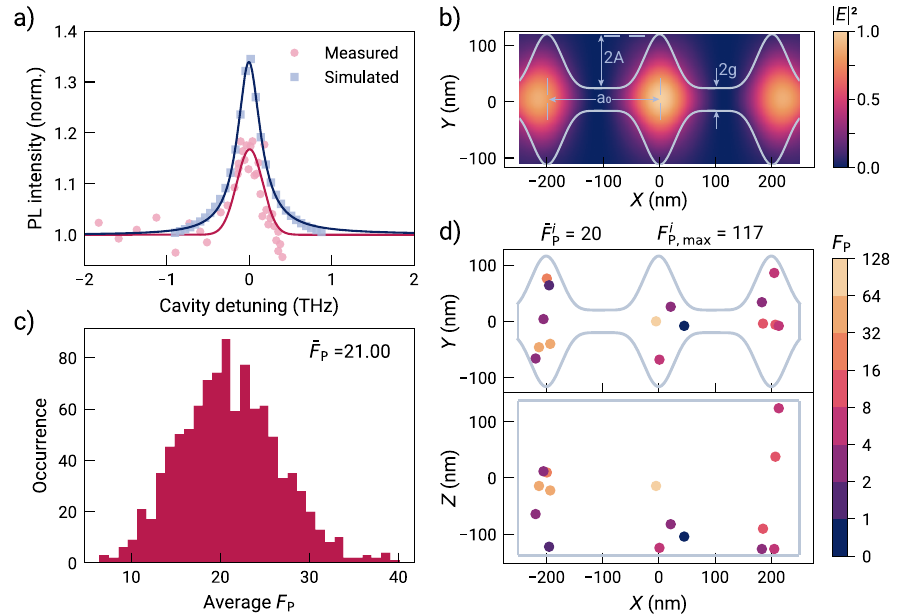}
    \caption{Numerical modelling of Purcell enhancement in the Sawfish cavity. a) Normalized PL intensity as a function of detuning between the cavity resonance and the NV ZPL for 15 NV centers in the confocal volume. Pink circles correspond to the measurement data and light blue squares originate from FEM simulations. The dark pink and dark blue solid lines indicate corresponding fits using a Voigt model. b) Normalized electric field intensity ($|E|^2$) of the simulated Sawfish cavity eigenmode visualized on a transverse plane passing through the center of the cavity, with the relevant geometric parameters indicated c) Histogram of the ensemble-averaged Purcell factor per emitter, obtained from the Monte-Carlo simulations. Each realization contains 15 cavity-embedded NV centers with randomly sampled positions and dipole orientations. d) Spatial distribution and color-coded Purcell factors of the 15 NVs for the Monte Carlo realization containing  the largest individual Purcell factor. The maximum ($F_\mathrm{P,max}^i$) and ensemble-averaged ($\bar{F}_\mathrm{P}^i$) Purcell factor for this realization ($i$) are specified. The color bar uses a base-2 logarithmic scaling. The grey lines mark the boundaries of the confocally addressed cavity volume.}
    \label{fig:fig3}
\end{figure*}

To further investigate the NV linewidth properties, we perform power-dependent PLE measurements on a single emitter (Fig.\,\ref{fig:fig1}d). The blue diamonds in Fig.\,\ref{fig:fig1}d show the inhomogeneous linewidth extracted from averaged PLE scans. Although averaging broadens the linewidth through spectral diffusion, the inhomogeneous linewidth narrows as the excitation power is reduced and approaches the lifetime limit of 13\,MHz (grey dashed line), determined independently from time-resolved fluorescence measurements (see Supplementary Materials). This indicates that spectral diffusion remains small even after averaging over many scans and several minutes. Within individual scans, spectral diffusion is quasi-static on millisecond timescales, such that the instantaneous lineshape is dominated by homogeneous processes \cite{orphal-kobin_optically_2023}. Using a Monte Carlo-based analysis \cite{orphal-kobin_retrieving_2025} to recover this information from the low signal-to-noise data (see Supplementary Materials), we extract the homogeneous linewidths (red circles in Fig.\,\ref{fig:fig1}d). These remain lifetime-limited up to excitation powers of 14\,nW.

\section{Purcell enhancement of optically coherent NVs} 
 The cavity-induced enhancement of NV emission is shown in Fig.\,\ref{fig:fig2}a. Through gas tuning, the Sawfish cavity resonance is spectrally overlapped with the NV ZPL \cite{mosor_scanning_2005, rugar_quantum_2021}. The cavity maintains a stable quality factor of 2400(515) over a range of 3\,THz (see Supplementary materials). The resulting cavity linewidth exceeds the inhomogeneous distribution of the NV ensemble by a factor of nearly 20. Therefore, when on resonance, all NVs in the confocal volume experience at least 99.8\% of the maximum cavity spectral response. Tuning the cavity into resonance increases the ZPL photoluminescence by 16.8(1.4)\% (Fig.\,\ref{fig:fig2}a, top), reflecting the cumulative enhancement of emission from multiple NVs.

We further compare the linewidths of individual NVs with the cavity tuned on and off resonance (Fig.\,\ref{fig:fig2}b–c). In both cases, the inhomogeneous linewidth remains around 30\,MHz at low excitation powers. At excitation powers of 40\,nW and above, the linewidth increases due to a combination of power broadening and excitation-induced spectral diffusion. Nevertheless, the inhomogeneous linewidth remains within 10 times the transform-limited linewidth. As each measurement is performed over different frequency ranges and initialized off-resonantly, we conclude that the measured narrow linewidths are not limited to a single emitter but are observed across multiple cavity-embedded NVs. Additionally, we compare the extracted homogeneous linewidths at 20\,nW (see Supplementary Materials). A small broadening is observed ($\Delta\Gamma <4\;\text{MHz}$) when the cavity overlaps with the NV ZPL. Considering this broadening to solely originate from a reduced excited-state lifetime, we estimate an effective Purcell factor of approximately 3.4(2.1) using a simple rate-equation model (see Supplementary Materials). However, the observed $\sim$17\% increase in PL is expected to be dominated by a small subset of NV centers that are favorably positioned and oriented within the cavity mode, while most emitters experience substantially weaker coupling.
\begin{figure*}[ht]
    \centering
    \includegraphics[]{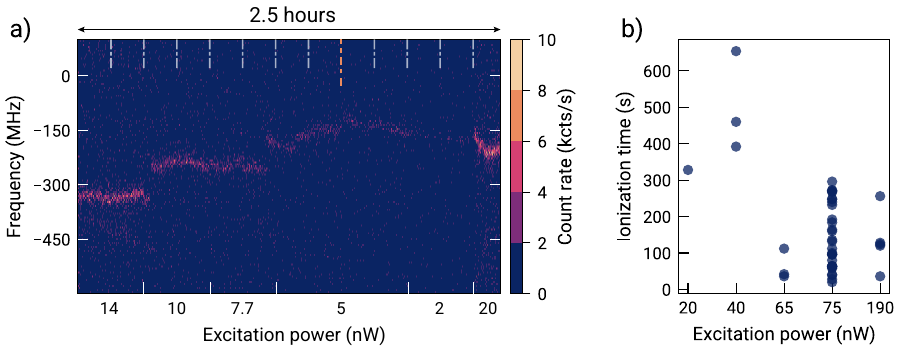}
    \caption{Charge-state stability under resonant excitation. a) Long-term PLE measurement of an NV center at varying excitation powers. The total measurement duration is approximately 2.5 hours. The frequency axis is given relative to 470.541 THz. The white dashed lines at the top of the panel indicate measurement interruptions of 1-2\,min, while the longer orange dashed line indicates a 1\,h interruption. The labels below the X-axis indicate the applied resonant excitation power during the measurement. b) Ionization times extracted from PLE measurements of individual NVs as a function of resonant excitation power.}
    \label{fig:fig4}
\end{figure*}

\section{Numerical modelling of Purcell enhancement}

Direct verification of strongly Purcell-enhanced individual emitters, for example through spectrally-resolved lifetime measurements, is hindered by a low signal-to-background ratio under pulsed resonant excitation. We therefore perform numerical simulations of a device-specific model of the Sawfish cavity interface using the FEM solver JCMsuite \cite{jcmwave} to determine the distribution of Purcell factors expected for randomly positioned and oriented NV centers. The simulated geometry is adjusted to reproduce the experimentally observed cavity quality factor. This yields a simulated $Q$-factor $Q_\mathrm{sim} = \text{2256}$, which agrees well with the measured value $Q_\mathrm{meas} = \text{2400(515)}$ (see Supplementary materials). For an optimally positioned and oriented emitter, this corresponds to a maximum Purcell factor ($F_\mathrm{P,ideal}$) of $\sim$210. To account for realistic emitter configurations, we introduce a reduction factor $\alpha_\mathrm{r}$, which depends on the projection of the local cavity field on the emitter dipole moment. The reduced Purcell factor is then given by $F_\mathrm{P,red} = \alpha_\mathrm{r} F_\mathrm{P,ideal}$.

To connect the simulated Purcell enhancement to the experimentally observed fluorescence increase, we model an ensemble of 15 dipoles with random positions and orientations within the addressed cavity volume. As the inhomogeneous broadening of the NV ensemble is small compared to the cavity linewidth, we neglect the spectral offsets between individual NVs and assign all dipoles the same emission frequency, while treating them as independent emitters. The simulated response reproduces the magnitude and overall lineshape of the experimentally observed enhancement, with only a small broadening and reduction in amplitude in the measured data (Fig.\,\ref{fig:fig3}a). To account for the small difference between the measured and simulated photoluminescence enhancement, we introduce a phenomenological scaling factor $\alpha_{\mathrm{s}}$, defined by the ratio of measured and simulated peak enhancements, and use it to normalize the simulated Purcell factors to the experimental system via $F_{\mathrm{P}} = \alpha_{\mathrm{s}}\cdot F_{\mathrm{P,red}}$. $\alpha_{\mathrm{s}}$ comes out to be approximately 0.88.

Finally, we perform Monte Carlo simulations to randomly place NVs in the confocal cavity volume, and use the calculated cavity eigenmode field to determine the statistical distribution of Purcell enhancement factors within the cavity-coupled NV ensemble (Fig.\,\ref{fig:fig3}b). In each iteration, we randomly sample dipole positions and polarizations for 15 emitters. Accounting for $\alpha_{\mathrm{r}}$ and $\alpha_{\mathrm{s}}$, the Monte Carlo simulations yield an average Purcell factor of $\bar{F}_\mathrm{P} = \text{21}$ and maximum values up to $F_\mathrm{P,max} = \text{117}$ (Fig.\,\ref{fig:fig3}c). The NV configuration yielding the maximum Purcell factor is shown in Fig.\,\ref{fig:fig3}d as an example, together with the corresponding Purcell factors. These results indicate that substantial Purcell enhancement can be achieved for randomly distributed emitters within the cavity.

The resulting Purcell enhancement increases the fraction of emission into the coherent ZPL. Without Purcell enhancement, the NV Debye-Waller factor $\xi_0$ \cite{debye_interferenz_1913} is roughly 3\%, representing the fraction of emission into the ZPL\cite{orphalkobin_coherent_2025}. The Purcell-enhanced coherent fraction is given by $\xi_\mathrm{P} = \sfrac{F_\mathrm{P}\xi_0}{(1+(F_\mathrm{P}-1)\xi_0)}$. For $F_\mathrm{P} = \text{117}$, this would yield $\xi_\mathrm{P} = \,\sim$78\% corresponding to a 26-fold increase in the ZPL emission fraction. This strong enhancement improves the suitability of cavity-integrated NV centers as optically coherent spin-photon interfaces for quantum networking.

\begin{figure}[ht]
    \centering
    \includegraphics[]{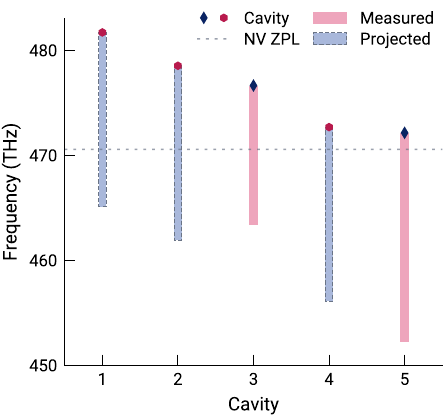}
    \caption{Multi-device spectral alignment of Sawfish cavities. Blue diamonds and red hexagons indicate the as-fabricated resonance frequencies of Sawfish cavities measured at cryogenic and room temperature, respectively. Pink bars represent the experimentally measured gas-tuning ranges for cavities 3 and 5. Blue bars with grey dashed outlines show the projected tuning ranges for the remaining cavities, defined as the average experimentally measured tuning range. The grey dashed horizontal line indicates the NV ZPL frequency}
    \label{fig:fig5}
\end{figure} 

\section{Charge-state stability}
Charge-state instability is a central challenge for NV centers in nanostructures, where off-resonant initialization can induce spectral jumps of several GHz \cite{faraon_coupling_2012}. To assess the long-term charge-state stability of the cavity-integrated NVs, we track the resonance frequency over 2.5 hours using repeated PLE scans (Fig.\,\ref{fig:fig4}a). After a single 520\,nm initialization pulse, only resonant excitation near 637\,nm is applied. During continuous resonant excitation, the ZPL exhibits slow spectral diffusion as well as occasional discrete jumps (Fig.\,\ref{fig:fig4}a), which we attribute to charge fluctuations in the local environment \cite{pieplow_quantum_2025}. The measurement includes interruptions of up to one hour, during which no optical excitation is applied. We find that the resonance frequency before and after these interruptions remains stable within approximately one natural linewidth, indicating that the NV does not undergo large spectral shifts during extended periods without optical excitation.

We further quantify ionization dynamics under varying resonant excitation powers (Fig.\,\ref{fig:fig4}b). For excitation powers below 50\,nW, ionization events are observed only on timescales exceeding 5\,minutes. Even for higher excitation powers the charge-state remains stable for several hundred seconds. This reduced susceptibility to ionization could be explained by recently proposed electron tunneling mechanisms for NV centers with many substitutional nitrogen atoms in its vicinity \cite{alkauskas_charge-state_2025, manson_spin_2026}.

\section{Multi-device spectral alignment}
A key requirement for scalable quantum photonic architectures is the spectral alignment of independently fabricated cavity–emitter systems. Using global in-situ gas tuning in combination with local laser ablation, we achieve an individual tuning range of about 18\,THz while maintaining the cavity Q-factor (Fig.\,\ref{fig:fig5}, cavity\,5 and Supplementary Materials). Using this tuning range as a reference, we identify 5 cavities on the same sample that can be tuned into resonance with the NV ZPL (red hexagons and blue diamonds in Fig.\,\ref{fig:fig5}), and experimentally tune a second cavity through the target resonance (Fig.\,\ref{fig:fig5}, cavity 3), demonstrating that individual cavity control is possible. Finally, using the average of the measured tuning ranges, we project the tuning range for the remaining cavities (blue-grey bars in Fig.\,\ref{fig:fig5}) under identical gas deposition conditions. Although the tuning mechanism itself is well established for global tuning, these results demonstrate individual spectral control of multiple spatially separated nanocavity–emitter interfaces on the same chip. Combined with the spectral multiplexing of multiple NV centers within a common cavity mode shown above, this provides a route towards architectures that scale both within and across nanophotonic devices.

\section{Conclusions}
We demonstrate that near lifetime-limited optical linewidths and long-lived charge-state stability can be preserved for charge-sensitive, spin-active emitters integrated into nanophotonic cavities. In contrast to the degradation of NV optical properties commonly observed after nanofabrication, the cavity-integrated NV centers investigated here retain their optical coherence over extended time scales while exhibiting Purcell-enhanced emission supported by numerical modelling. These results overcome the longstanding challenge of integrating optically coherent NV centers with nanophotonic cavities. Within a common cavity mode, the preserved optical coherence enables multiple spectrally distinct NV centers to be coupled and individually addressed, establishing the frequency multiplexing of cavity-integrated NV centers. In addition, the independent spectral control of spatially separated cavity-emitter devices on the same chip provides a complementary route towards scaling cavity-integrated quantum network nodes. A next step is to improve photon extraction, for example through evanescent coupling to tapered optical fibers \cite{burek_fiber-coupled_2017}. Together, these developments could enable scalable cavity-integrated quantum network nodes based on NV centers.

%% file: methods.tex
\section{Methods}

\subsection{Sample fabrication}
The diamond substrate used in this work was a $3\times3\times0.25~\mathrm{mm}^3$ type-IIa optical-grade diamond supplied by Element Six, with a specified substitutional nitrogen concentration below $1\,\mathrm{ppm}$ \cite{E6_website}. The investigated NV centers were native defects formed during diamond growth; no implantation or irradiation was used to create additional NV centers. Suspended Sawfish photonic crystal cavities with a nominal thickness of $275\,\mathrm{nm}$ were fabricated using quasi-isotropic under-etching \cite{mouradian_rectangular_2017}. The complete fabrication process and the Sawfish cavity geometry are described in Refs.~\cite{pregnolato_fabrication_2024,bopp_sawfish_2024}.

\subsection{Cryogenic optical characterization}
All optical measurements were performed at a sample temperature of $5\,\mathrm{K}$. Cavity resonances were characterized by PL spectroscopy under non-resonant excitation and fitted with Lorentzian line shapes to determine their resonance wavelengths and quality factors. Resonant PLE spectroscopy was performed by repeatedly scanning a narrow-linewidth laser across the NV ZPL while detecting fluorescence in the phonon sideband (PSB). The excitation powers quoted throughout the manuscript refer to the power incident on the sample.

Cavity-induced fluorescence enhancement was determined by monitoring the spectrally integrated NV emission while tuning the cavity resonance through the NV ZPL. The measured intensity was normalized to the PSB fluorescence level. Because no lifetime reduction was resolved experimentally, the measured quantity is reported as a fluorescence enhancement and compared with the enhancement expected from the numerical model.

\subsection{PLE spectroscopy and linewidth analysis}
Each linewidth dataset consisted of 150-400 consecutive scans of the resonant laser across the NV ZPL. Individual scans lasted approximately $1\,\mathrm{s}$. The laser remained resonant with a given transition for only a few milliseconds during each scan, corresponding to an excitation duty cycle below $1\%$. Scan-averaged spectra were fitted with a Voigt profile and a constant background. The linewidth extracted from the averaged spectrum is referred to as the inhomogeneous linewidth and includes spectral fluctuations occurring over the full acquisition time.

At low excitation powers, the small number of photons detected during an individual scan does not fully sample the resonance line shape and biases direct single-scan fits towards smaller linewidths. The homogeneous linewidth was therefore inferred by comparing the measured single-scan photon-counting statistics with Monte Carlo simulations generated for a range of trial linewidths and mean photon numbers, following the method described in Ref.~\cite{orphal-kobin_retrieving_2025}. Details of the photon-counting model, inference procedure and validation are provided in the Supplementary Information.

For charge-stability measurements, non-resonant initialization was applied only before the first PLE scan. Repeated resonant scans were subsequently acquired without further charge-state reinitialization. Charge-state stability was inferred from the continued observation of the same optical transition, while its fitted center frequency was used to quantify spectral shifts over time.

\subsection{Gas tuning and multi-device alignment}
Ultra-pure nitrogen gas (see Supplementary Materials) was introduced into the cryogenic sample chamber through a fine-dosing valve and a capillary tube. Condensation of nitrogen onto the sample surface increased the refractive index surrounding the cavities and red-shifted their resonances. Spatially selective blue-tuning was achieved by locally desorbing the condensed nitrogen using approximately $2\,\mathrm{mW}$ of continuous-wave $520\,\mathrm{nm}$ laser light.

Global gas-condensation and local laser-desorption steps were applied while monitoring the cavity spectra. This procedure enabled spatially separated cavities on the same chip to be tuned individually until their resonances overlapped with selected NV zero-phonon transitions.

\subsection{Numerical simulations}
Finite-element simulations of the Sawfish cavity and the collected NV fluorescence were performed using JCMsuite \cite{jcmwave}. The electromagnetic response was calculated using a frequency-domain scattering formulation for $15$ fluorescing NV centers distributed within the cavity excitation volume. The NV positions and dipole orientations were sampled according to the spatial and crystallographic distributions described in the Supplementary Materials. Because the NV centers emit independently, the collected powers from the individual dipole sources were summed incoherently.

The electromagnetic field was evaluated at a far-field plane located more than $1.3\,\mu\mathrm{m}$ above the cavity surface, where the contribution of the evanescent cavity field was negligible. The collected signal was obtained from the angular spectrum of the radiated field by integrating over transverse wavevectors satisfying
$$
k_{\parallel}\leq k_0\mathrm{NA},
$$
where $k_{\parallel}$ is the magnitude of the wavevector component parallel to the collection plane, $k_0$ is the free-space wavevector and $\mathrm{NA}$ is the numerical aperture of the collection objective. The simulated cavity-induced enhancement was calculated relative to the corresponding off-resonant emission and compared directly with the experimentally measured fluorescence enhancement. Full simulation parameters and convergence tests are provided in the Supplementary Materials.

%% file: supplement_text.tex
\section{Supplementary Note 1: Experimental platform}

\subsection{Diamond sample and Sawfish cavity fabrication}
Experiments were performed on a 3$\times$3$\times$0.25\,mm type-IIa optical-grade diamond substrate supplied by Element Six. The manufacturer specifies a substitutional nitrogen concentration below 1\,ppm \cite{E6_website}. The investigated NV centers were native defects formed during crystal growth; no implantation or irradiation was used to create additional NV centers.

The polished surface of the substrate was oriented along $[\,\text{001}\,]$, and the longitudinal axes of the devices were aligned along $[\,\text{100}\,]$. Suspended Sawfish photonic crystal cavities with a nominal thickness of 275\,nm were fabricated using quasi-isotropic oxygen-plasma under-etching \cite{mouradian_rectangular_2017}. The complete fabrication sequence and the nominal cavity design are described in \cite{pregnolato_fabrication_2024,bopp_sawfish_2024}. A representative completed device is shown in Supplementary Fig.\,\ref{fig:sawfish_sem}.
\begin{figure}[h]
    \centering
    \includegraphics[]{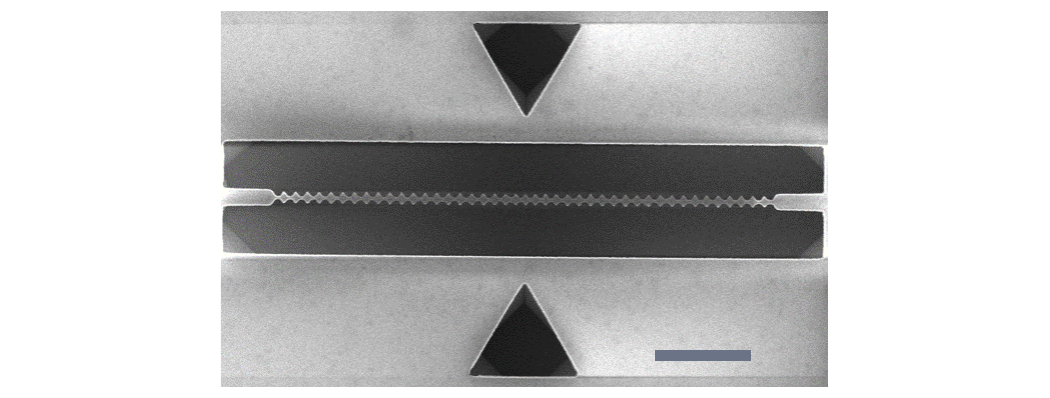}
    \caption{SEM of a representative suspended symmetric Sawfish cavity. The scale bar is 2\,$\mu$m}
    \label{fig:sawfish_sem}
\end{figure}

\subsection{Cryogenic confocal microscope}
The sample was mounted in a closed-cycle helium cryostat (attoDRY800, attocube systems) and maintained at a temperature of approximately 5\,K. Excitation and collection were performed normal to the diamond surface through a cryogenic objective with a numerical aperture of 0.82 (LT-APO/VISIR/0.82, attocube systems).

Off-resonant excitation was provided by a 520\,nm laser (Oxxius), whereas resonant excitation of the NV zero-phonon line (ZPL) was performed using a narrow-linewidth 637\,nm external-cavity diode laser (DL pro, TOPTICA Photonics). Unless stated otherwise, all powers refer to the optical power incident on the sample.

The emitted fluorescence was collected through the excitation objective and spectrally restricted to the range between approximately 550 and 800\,nm using long-pass and short-pass filters.
For measurements requiring simultaneous detection of the ZPL and the phonon sideband (PSB), a 649\,nm long-pass dichroic separated the two spectral regions. The ZPL channel was coupled to a grating spectrometer (SpectraPro HRS-500, Princeton Instruments), while the PSB channel was coupled to a fiber-coupled avalanche photodiode (SPCM-AQRH, Excelitas Technologies). The 649\,nm spectral-separation stage was used for the measurements shown in Fig.\,5 and removed for measurements shown in Fig.\,2 of the main text. This difference accounts for the different detected spectral intensities in Supplementary Figs.\,\ref{fig:exp_schematic}b and \ref{fig:spec-norm}.
\begin{figure}
    \centering
    \includegraphics[width=0.8\linewidth]{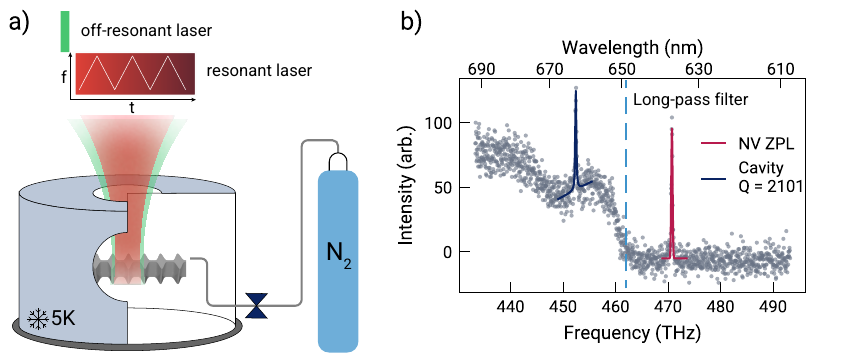}
    \caption{a) Simplified schematic of the experimental setup. The light grey cylindrical box represents the sample chamber of the cryostat. The green and red light cones represent the path of the respective excitation lasers that are focused onto the center of the Sawfish cavity. Depicted on top is the PLE measurement protocol. It starts off with a green laser pulse for initialization. Afterwards, only the red  laser frequency is scanned across the NV ZPL. The gas-feed line is pictured on the right. Nitrogen gas sitting in a reservoir ($N_2$ bottle) passes through a capillary tube (grey line) before diffusing onto the Sawfish cavity. The amount of gas going into the cryostat is controlled using a fine-dosing valve (dark blue hourglass shape). b) A fluorescence spectrum of the Sawfish cavity demonstrating the tuning range of the cavity referred to in the main text. Grey dots are the measured data-points after background subtraction. The blue dashed line represents the edge of the 649\,nm dichroic filter. The red and blue lines denote the NV ZPL and the cavity resonance respectively.}
    \label{fig:exp_schematic}
\end{figure}

\subsection{Resonant excitation, frequency control and data acquisition}
The resonant excitation power was stabilized using a fiber-coupled acousto-optic modulator (Fiber-Q, Gooch \& Housego) and a photodiode in a closed feedback loop implemented on a Red Pitaya platform. The laser frequency was monitored using a WS7 wavemeter (HighFinesse).

Laser scanning, photon counting and synchronization were controlled using a National Instruments USB-6353 data-acquisition device and the Qudi control framework \cite{binder_qudi_2017}. Gas dosing and laser-induced desorption were controlled using DynExp \cite{bopp_dynexphighly_2024}.

\section{Supplementary Note 2: Number of optically addressed NV centers}

\subsection{Estimate from nominal material parameters}

The manufacturer specifies a substitutional nitrogen concentration $c_{\mathrm{N}}<1\,\mathrm{ppm}$. A representative ratio of approximately $300{:}1$ between substitutional nitrogen and negatively charged NV
centers has been reported for nitrogen-doped CVD diamond \cite{Edmonds_production_2012}. Within this approximate conversion model, the NV density is
$$
\rho_{\mathrm{NV}}^{(\mathrm{mat})} \simeq \frac{c_{\mathrm{N}}n_{\mathrm{C}}}{300},
$$
where $n_{\mathrm{C}}=1.76\times10^{23}\,\mathrm{cm}^{-3}$ is the atomic number density of diamond.

The excitation volume was calculated from the overlap of the diffraction-limited excitation spot with the in-plane geometry of the Sawfish cavity,
$$
V_{\mathrm{exc}} = t\iint \chi_{\mathrm{cav}}(x,y) \chi_{\mathrm{spot}}(x,y) \,\mathrm{d}x\,\mathrm{d}y,
$$

where $t$=275\,nm is the device thickness and $\chi_{\mathrm{cav}}$ and $\chi_{\mathrm{spot}}$ describe the cavity geometry and excitation region, respectively. We use an effective excitation-spot diameter of $494\,\mathrm{nm}$, corresponding to the diffraction-limited Gaussian diameter for $\lambda$=637\,nm and $\mathrm{NA}=0.82$. Integration over the cavity geometry gives
$$
V_{\mathrm{exc}}=0.01514\,\mu\mathrm{m}^{3}.
$$
The nominal material parameters therefore become
$$
N_{\mathrm{NV}}^{(\mathrm{mat})} \simeq \rho_{\mathrm{NV}}^{(\mathrm{mat})}V_{\mathrm{exc}} <8.8.
$$

This value provides an order-of-magnitude estimate rather than a strict upper bound, because the manufacturer specification does not determine the local nitrogen concentration in the nanostructured region and the N-to-NV conversion ratio is sample-dependent.

\subsection{Estimate from PLE spectroscopy}
\begin{figure}
    \centering
    \includegraphics[width=0.85\linewidth]{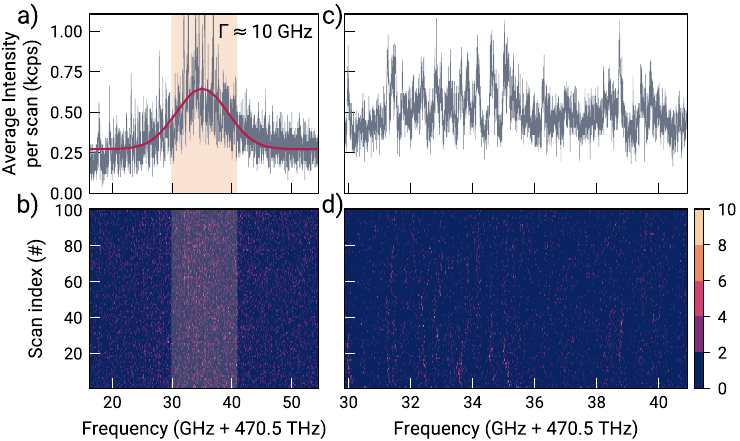}
    \caption{Estimating NV number. a) PLE spectrum at 75\,nW resonant excitation power. A scan range of 38\,GHz is used to fully address the NV distribution. The inhomogeneous distribution of NV ZPL frequencies has a Gaussian lineshape with an FWHM of around 10 GHz. b) The full measurement of the data shown in a) comprises of 100 PLE scans that are each visualized here horizontally and stacked vertically. The highlighted area in both a) and b) marks the scan range in c) and d). c-d) Analogous measurement to a-b) with a narrower scan range of 11 GHz at the same excitation power. Individual NVs can be counted using the peaks in c) and manually correlated with the time traces in d)}
    \label{fig:nv_num_est}
\end{figure}
A broad-range PLE measurement spanning 38\,GHz was used to determine the inhomogeneous distribution of NV ZPL transition frequencies (Supplementary Fig.\,\ref{fig:nv_num_est}a-b). The scan-averaged spectral envelope was fitted with a Gaussian function with an FWHM of approximately 10\,GHz, corresponding to
$$
\sigma = \frac{\mathrm{FWHM}}{2\sqrt{2\ln 2}} \simeq4.25\,\mathrm{GHz}.
$$

Higher-resolution measurements were subsequently acquired over an 11\,GHz interval centered on the distribution (Supplementary Fig.\,\ref{fig:nv_num_est}c,d). Individual resonances were identified from peaks in the integrated spectrum and verified by following their trajectories in the scan-resolved PLE maps. Across 6 measurements, a mean of $14.5$ resonances was identified within this interval.

For a centered interval of width $\Delta\nu=11\,\mathrm{GHz}$, the fraction of the fitted Gaussian distribution contained within the measurement window is
$$
f_{\mathrm{win}} = \operatorname{erf} \left( \frac{\Delta\nu}{2\sqrt{2}\sigma} \right) \simeq0.805. $$

The corresponding number of optical resonances across the complete inhomogeneous distribution is therefore

$$
N_{\mathrm{tr}} = \frac{14.5}{f_{\mathrm{win}}} \simeq18.
$$

The scan-resolved data permit individual resonance trajectories to be identified, but their signal-to-noise ratio is insufficient to unambiguously determine whether some pairs of transitions originate from the two orbital branches of the same NV center. Previous measurements of NVs in comparable substrates and inside nanostructures indicate that frequently only one orbital transition is observable. We therefore expect the number of resolved transitions to be comparable to the number of optically observable NV
centers, while allowing for occasional double counting of centers for which both branches are visible.

Conversely, weak or overlapping transitions may remain unresolved, particularly toward the tails of the inhomogeneous distribution. The PLE measurements were performed under resonant excitation following charge-state initialization, without simultaneous off-resonant repumping, reducing repump-induced spectral reconfiguration during the scans.

Taking the directly observed mean number of resonances as the central estimate, and allowing for the extrapolation to the unmeasured spectral tails, ambiguous transition assignments, and peak-identification uncertainty, we estimate
$$
N_{\mathrm{NV}}=15(4).
$$
The quoted uncertainty is a conservative estimate reflecting these systematic ambiguities rather than the statistical uncertainty of the mean alone.

\section{Supplementary Note 3: PLE spectroscopy and optical coherence}

\subsection{PLE acquisition protocol}
Each PLE measurement consisted of several hundred consecutive scans of the resonant laser across the selected NV transitions.
A single scan lasted approximately 1\,s, with the laser remaining within one linewidth of a given transition for only a few milliseconds.
Before the first scan, a 520\,nm pulse with a power of approximately 200\,$\mu$W was applied to initialize the NV charge state. No further non-resonant illumination was applied during the measurement. Fluorescence in the PSB was recorded using the APD while the resonant-laser frequency was scanned across the ZPL.

\subsection{Monte Carlo inference of homogeneous linewidths}
For PLE measurements with low count rates (Supplementary Fig.\,\ref{fig:mc_lw_est}a) and consequently low signal-to-noise ratio (SNR, yellow diamonds in Supplementary Fig.\,\ref{fig:mc_lw_est}b), direct fits to individual scans preferentially recover narrow peaks and therefore underestimate the underlying single-scan linewidth (blue diamonds in Supplementary Fig.\,\ref{fig:mc_lw_est}b). We infer the homogeneous linewidth by reproducing the complete single-scan fitting procedure using Monte Carlo-generated photon-counting data, following Ref.\,\cite{orphal-kobin_retrieving_2025} (red circles in Supplementary Fig.\,\ref{fig:mc_lw_est}b).

For each trial linewidth $\Gamma$ and mean signal-photon number $\bar{N}_{\mathrm{s}}$, the total number of signal photons in each simulated scan was drawn from a truncated normal distribution with the measured mean and standard deviation. Signal-photon frequencies were sampled from a Lorentzian line shape of FWHM $\Gamma$ within a 150\,MHz analysis window. Background counts were drawn from a Poisson distribution with a mean determined from the corresponding experimental dataset and distributed uniformly over the same window.

Each simulated scan was processed and fitted using the same procedure as the experimental data. The resulting fitted linewidths were compiled into a histogram (Supplementary Fig.\,\ref{fig:mc_lw_est}c). For each pair $(\Gamma,\bar{N}_{\mathrm{s}})$, $20,000$ simulated scans were generated, and the simulated histogram was compared with the experimental histogram and the best fitting parameter pair was computed using a $\chi^2$ test (Supplementary Fig.\,\ref{fig:mc_lw_est}d).

The best-fitting homogeneous linewidth and signal-photon number were defined by the global minimum of the $\chi^2$ surface. Confidence intervals for each parameter were calculated by profiling over the other parameter. The $3\sigma$ interval was defined by
$$
\chi^2-\chi^2_{\min}=9.
$$
The resulting intervals are one-dimensional profile intervals and should not be interpreted as a joint two-parameter confidence contour.
\begin{figure}
    \centering
    \includegraphics[width=0.9\linewidth]{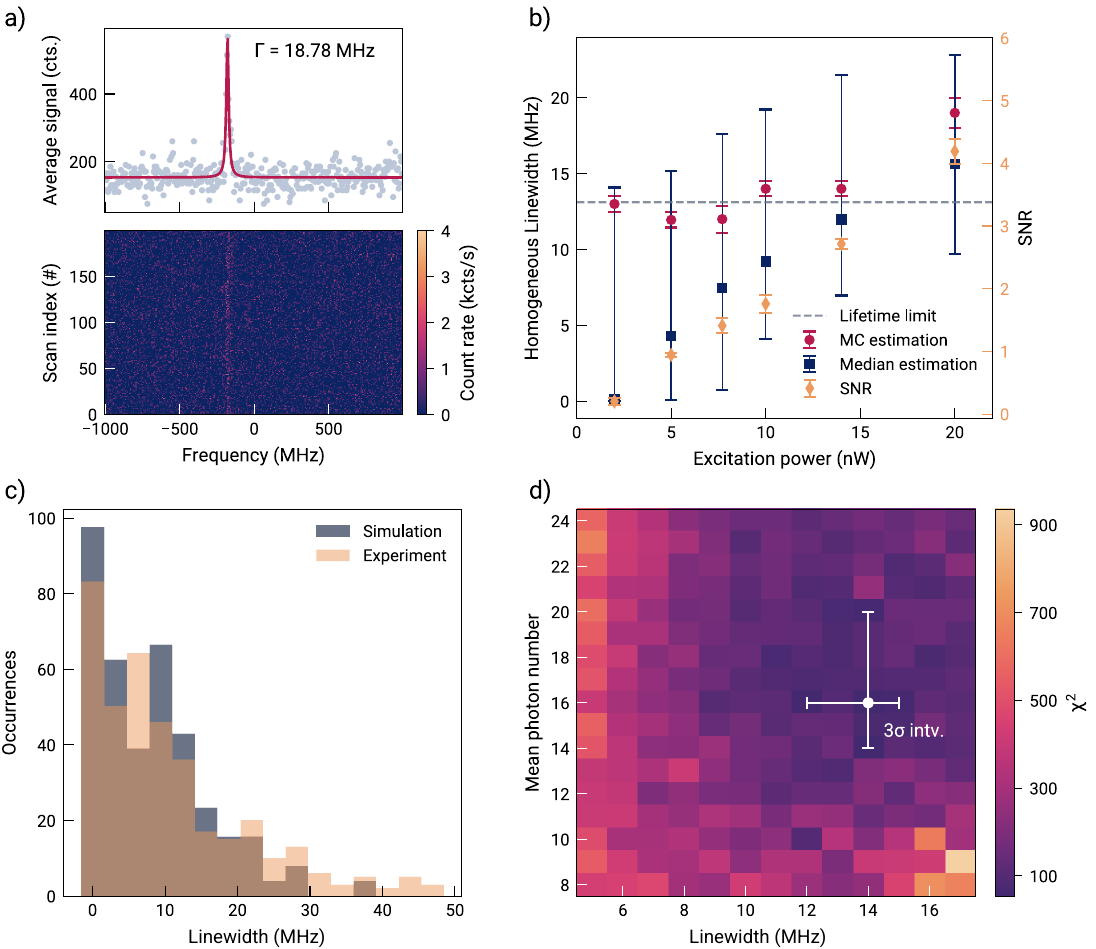}
    \caption{Monte-Carlo linewidth estimation. a) PLE measurement at 2\,nW excitation power. the bottom panel shows the time-trace of the NV ZPL, from repeated scanning of the resonant laser. The top panel shows the integrated signal (light-gray circles) and a Voigt fit (solid red line) revealing an inhomogeneous linewidth of about 19\,MHz. The frequency axis is relative to 470.541\,THz. b) Power-dependent homogeneous linewidth and signal-to-noise ratio (SNR) extracted from PLE measurements of the same NV as in a). The red circles represent values extracted using the Monte-Carlo method, with the error-bars indicating the $\text{3}\sigma$ confidence interval. The blue squares are median values of the distribution of single-scan linewidths for each excitation power, with the error-bars representing the inter-quartile range. The yellow diamonds represent the measured SNR for each excitation power, with the error-bars indicating the measurement uncertainty. the gray dashed line represents a reference value for the lifetime-limited linewidth. c-d) Example results of the Monte-Carlo estimation for a resonant excitation power 10\,nW. c) Single-scan linewidth histograms computed from the simulated (dark blue) and experimental (light yellow) data. The brown color represents regions where the two histograms overlap. d) Distribution of the $\chi^2$ value as a function of the linewidth and mean photon number in the Monte-Carlo simulations. The white dot represents the best fit and corresponds to the histogram shown in c). The error-bars indicate the 3$\sigma$ confidence interval. The confidence intervals for linewidth and mean photon number are calculated independent of each other.}
    \label{fig:mc_lw_est}
\end{figure}

\section{Supplementary Note 4: Cavity coupling and fluorescence enhancement}

\subsection{Cavity resonance and quality-factor extraction}
\begin{figure}[h!]
    \centering
    \includegraphics[width=0.95\linewidth]{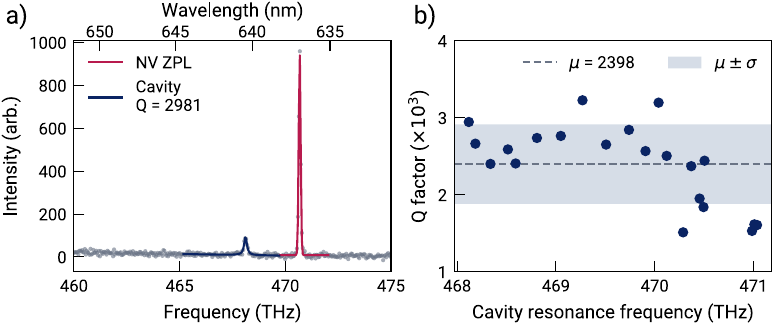}
    \caption{Measuring the cavity quality factor during the resonance-tuning process. a) Representative fluorescence spectrum at the beginning of the resonance tuning process. Gray dots are the measured signal. The blue and red lines represent the cavity resonance and NV ZPL respectively. b) Tracking the cavity Q-factor as a function of its resonance frequency. The grey dashed line is the average Q-factor and the shaded region represents the standard deviation of the measured Q-factors.}
    \label{fig:cavity_q}
\end{figure}

Cavity spectra were recorded under 520\,nm excitation using the grating spectrometer. After subtraction of a spectrally smooth background, the cavity resonance was fitted with a Lorentzian line shape. The cavity quality factor was calculated as $Q=\sfrac{\nu_{\mathrm{c}}}{\Delta\nu_{\mathrm{c}}},$
where $\nu_{\mathrm{c}}$ and $\Delta\nu_{\mathrm{c}}$ are the fitted resonance frequency and
FWHM, respectively (Supplementary Fig.\,\ref{fig:cavity_q}a). 

\subsection{Normalization of the PL spectra}
\begin{figure}
    \centering
    \includegraphics[width=0.5\linewidth]{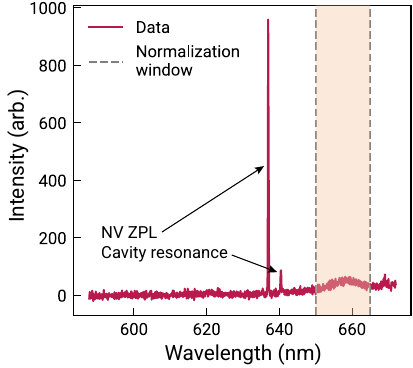}
    \caption{Example spectrum indicating the normalization used in Fig.4 in the main text. The red signal is a fluorescence measurement from the sawfish cavity. The yellow shaded region represents the spectral range in which fluorescence counts are summed and used as the normalization factor for the entire spectrum}
    \label{fig:spec-norm}
\end{figure}

Gas deposition and local laser heating caused small variations in the sample position and excitation efficiency during the tuning sequence. Each recorded spectrum was therefore normalized using the integrated
counts in a PSB interval between 650--665\,nm. This interval was spectrally separated from the cavity resonance and was assumed not to be modified by the cavity tuning. It therefore provided a reference for common-mode variations in excitation and collection efficiency (Supplementary Fig.\,\ref{fig:spec-norm}).

After background subtraction, the normalization factor was calculated as
$$
I_{\mathrm{ref}} = \int_{\lambda_1}^{\lambda_2} I(\lambda)\,\mathrm{d}\lambda,
$$
and each spectrum was divided by $I_{\mathrm{ref}}$. The selected integration window is indicated in Supplementary Fig.\,\ref{fig:spec-norm}.

\subsection{Definition of the cavity-induced fluorescence enhancement}
For each cavity detuning, the normalized ZPL signal was calculated as
$$
R(\Delta) = \frac{I_{\mathrm{ZPL}}(\Delta)}{I_{\mathrm{PSB}}(\Delta)},
$$

where $I_{\mathrm{ZPL}}$ is the integrated intensity in the spectral window $\pm$175\,GHz around the ZPL center frequency, $\Delta$ is the cavity detuning and $I_{\mathrm{PSB}}$ is the reference intensity defined above. The cavity-induced fluorescence enhancement was then calculated relative to the off-resonant value,
$$
\mathcal{E}(\Delta) = \frac{R(\Delta)}{\langle R\rangle_{\mathrm{off}}},
$$

where the off-resonant reference was obtained for cavity detunings larger than 1\,THz from the NV ZPL. Uncertainties were obtained from the fit covariance.

\subsection{Ensemble lifetime measurements}
\begin{figure}
    \centering
    \includegraphics[width=0.6\linewidth]{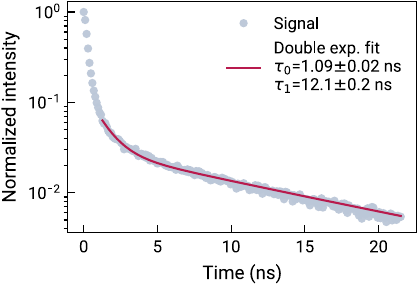}
    \caption{Time-resolved fluorescence measurement from cavity-embedded NVs. Gray circles represent the measured signal and the red line is a fit to a double-exponential model.}
    \label{fig:lft_ref}
\end{figure}
Time-resolved fluorescence of the cavity-embedded NVs was recorded under pulsed 520\,nm excitation. Optical pulses with a length of about 1\,ns and a repetition period of 102\,ns were generated by a supercontinuum laser source (SuperK Fianium with a SuperK Varia filter, NKT Photonics). Photon arrival times were recorded using a fiber-coupled APD and a time tagger (QuTools QuTAG-MC).

The fluorescence decay curve was fitted with a double-exponential model after the excitation pulse fell below the signal level (Supplementary Fig.\,\ref{fig:lft_ref}). The fast decay originates from the residual excitation pulse intensity whereas the slow decay (12.1\,ns) gives the excited state lifetime of the NV ensemble. This corresponds to a lifetime-limited linewidth of about 13.11\,MHz.

\subsection{Estimate of the Purcell enhancement from homogeneous linewidths}
\begin{figure}
    \centering
    \includegraphics[]{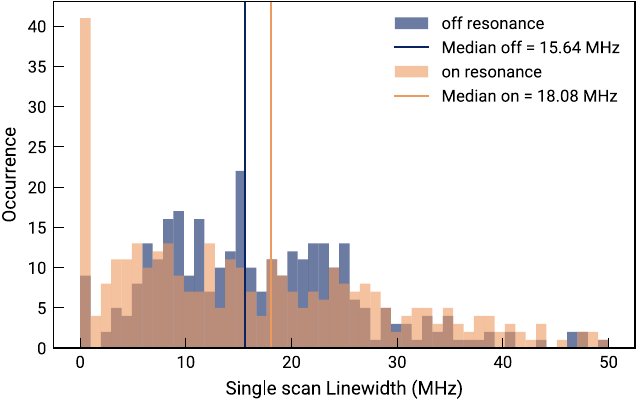}
    \caption{Histogram of the single scan linewidths of single NVs measured at an excitation power of 20\,nW when the cavity is on and off resonance (yellow and blue respectively). The yellow and blue solid vertical lines mark the median homogeneous linewidth for the on and off resonant case respectively. The total sample size os 400 PLE scans each.}
    \label{fig:hom_lw_hist}
\end{figure}

As shown in \cite{orphal-kobin_retrieving_2025}, at sufficiently high photon count rates, the median single-scan linewidth provides a suitable estimator for the homogeneous linewidth. We therefore compare homogeneous linewidths measured at a resonant excitation power of 20\,nW with the cavity tuned on and off resonance with the NV ZPL. At this excitation power, the NVs are well below optical saturation and power broadening is negligible. A representative comparison of the corresponding single-scan linewidth distributions is shown in Fig.\,\ref{fig:hom_lw_hist}.

The cavity-induced increase in the radiative decay rate is expected to broaden the homogeneous optical linewidth. Let $\gamma_0$ denote the total decay rate of the NV excited state, with contributions $\gamma_\mathrm{ZPL}$ and $\gamma_\mathrm{PSB}$ from emission into the ZPL and PSB, respectively,
$$\gamma_0=\gamma_\mathrm{ZPL}+\gamma_\mathrm{PSB}.$$
Using the conventional definition of the Purcell factor employed throughout this work, the cavity introduces an additional ZPL decay rate
$$\gamma_\mathrm{cav}=F_\mathrm{P}\gamma_\mathrm{ZPL},$$
such that the total decay rate on cavity resonance becomes
$$\gamma_\mathrm{ON}=\gamma_0+F_\mathrm{P}\gamma_\mathrm{ZPL}.$$
Since the measurements are performed well below optical saturation, the homogeneous linewidth is determined by the lifetime-limited contribution together with any residual pure-dephasing contribution. Assuming that the latter is comparable for the NVs measured on and off cavity resonance, the cavity-induced linewidth change is
$$\Delta LW=LW_\mathrm{ON}-LW_\mathrm{OFF}=\frac{F_\mathrm{P}\gamma_\mathrm{ZPL}}{2\pi},$$
which gives
$$F_\mathrm{P}=\frac{2\pi\Delta LW}{\gamma_\mathrm{ZPL}}.$$
The ZPL decay rate is estimated from the measured excited-state lifetime $\tau$ (Fig.\,\ref{fig:lft_ref}) and a Debye--Waller factor of 0.03 according to
$$\gamma_\mathrm{ZPL}=\frac{0.03}{\tau},$$
yielding $\gamma_\mathrm{ZPL}=2.2\,\mathrm{MHz}$. The median homogeneous linewidth measured with the cavity off resonance was 15.64\,MHz and is used as the reference linewidth. The Purcell factors inferred from the on-resonant linewidth measurements are summarized in Table\,\ref{tab:hom_lw_to_fp}.

\begin{table}[htbp]
    \centering
    \caption{Estimated Purcell factors inferred from the increase in homogeneous linewidth relative to the off-resonant reference linewidth of 15.64\,MHz.}
    \begin{tabular}{|c|c|}
        \hline
        $LW_\mathrm{ON}$ (MHz) & $F_\mathrm{P}$ \\
        \hline
        18.08 & 6.90 \\
        16.49 & 2.39 \\
        16.10 & 1.31 \\
        16.75 & 3.13 \\
        \hline
    \end{tabular}
    \label{tab:hom_lw_to_fp}
\end{table}

Only one off-resonant reference measurement is available, whereas the on-resonant linewidth measurements contain contributions from at least two different NV centers that cannot be individually identified. The extracted Purcell factors should therefore be regarded as estimates rather than emitter-specific measurements. The analysis assumes that the measured off-resonant homogeneous linewidth is representative of the intrinsic linewidths of the NVs contributing to the on-resonant measurements.

\section{Supplementary Note 5: Numerical modeling}

\subsection{Sawfish cavity eigenmode}
\begin{figure}
    \centering
    \includegraphics[width=0.95\linewidth]{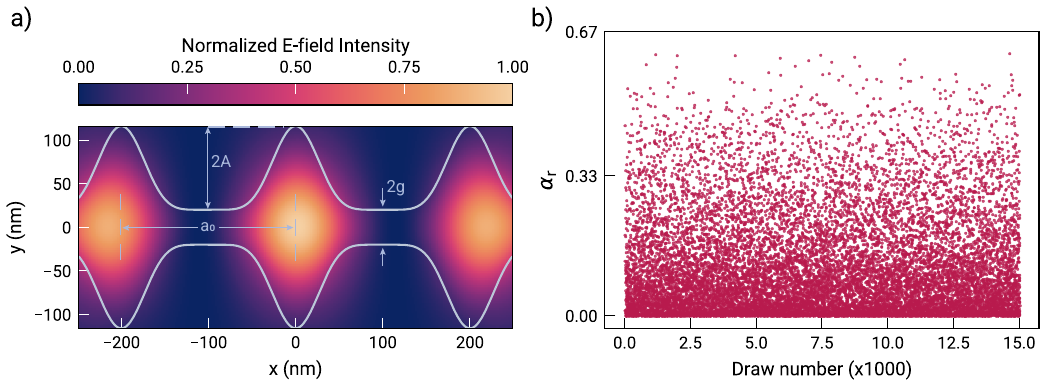}
    \caption{Numerical modeling of the Sawfish cavity. a)  Electric field intensity distribution of the fundamental cavity mode in a transverse slice at the center of the cavity. The grey lines mark the boundaries of the Sawfish structure. b) Distribution of the reduction factor $\alpha_\mathrm{r}$ of the ideal Purcell enhancement ($F_\mathrm{P,ideal}$) due to spatial and dipole mismatch of the cavity-embedded NVs. The X axis indicates individual iterations of Monte-Carlo simulations varying the position and orientation of an NV inside the Sawfish cavity.}
    \label{fig:fe_sim}
\end{figure}

Finite-element simulations were performed using JCMsuite \cite{jcmwave}. The fundamental cavity mode was calculated using the eigenmode solver for the device geometry listed in Supplementary Table\,\ref{tab:cav_params}. The simulated geometry was adjusted within the fabrication uncertainty and simulation constraints to reproduce the experimentally measured cavity wavelength and quality factor (Supplementary Fig.\,\ref{fig:fe_sim}a).
\begin{table}[h]
    \centering
    \caption{Cavity geometry parameters used for all calculations. Parameters are determined such that the simulated cavity resonance and quality factor match the measured values.}
    \vspace{4 pt}
    \begin{tabular}{|c|c|}
        \hline
        Cavity parameter & Value \\
        \hline
        A & 48 nm \\
        $a_0$ & 200 nm \\
        g & 20 nm \\
        T & 275 nm \\
        N & 11 \\
        \hline
    \end{tabular}
    \label{tab:cav_params}
\end{table}

The calculated mode had a resonance wavelength of 636.36\,nm, a quality factor of 2256, and a mode volume of 0.81 in units of $(\lambda/n)^3$, where $\lambda$ is the free-space transition wavelength and $n$ is the refractive index of diamond. The electric-field intensity in the central plane of the cavity is shown in Supplementary Fig.\,\ref{fig:fe_sim}.

\subsection{Ideal Purcell factor}
For a resonant dipole placed at the electric-field maximum and aligned with the local field polarization, the ideal Purcell factor is
$$
F_{\mathrm{P,ideal}} = \frac{3}{4\pi^2} \frac{Q}{V} \left(\frac{\lambda}{n}\right)^3.
$$
Here, $Q$ and $V$ are the simulated cavity quality factor and mode volume. This $F_{\mathrm{P,ideal}}$ refers to the Purcell enhancement of the ZPL transition only.

\subsection{Spatial and dipole-orientation distribution}
For an emitter at position $\mathbf{r}_i$ with optical dipole $\mathbf{d}_i$, the local reduction of the ideal Purcell factor was calculated as

$$
\alpha_\mathrm{r} = \frac{\left| \mathbf{d}_i\cdot\mathbf{E}(\mathbf{r}_i) \right|^2}
{|\mathbf{d}_i|^2 \max_{\mathbf r}|\mathbf{E}(\mathbf r)|^2},
$$
such that
$$
F_{\mathrm{P,red}} = \alpha_\mathrm{r} F_\mathrm{P,ideal}.
$$
This expression includes both the reduced field intensity away from the mode maximum and the mismatch between the dipole and local field polarizations. Each NV axis was sampled from the crystallographically allowed $\langle111\rangle$ orientations. The optical dipoles were constrained to the plane perpendicular to the corresponding NV axis \cite{Maze_properties_2011}. Since the polarization of the optical dipoles depends on local strain close to the NV location and cannot be assigned deterministically, the dipoles are assigned random polarizations in the plane they are confined in.

Each ensemble realization contained 15 NV centers. Their positions were sampled from a uniform distribution in the physical cavity volume bounded by the diameter of the focused Gaussian profile of the excitation laser. Their NV axes and optical dipoles were sampled as described above. A total of 1000 independent ensemble realizations were generated. The distribution of $\alpha_\mathrm{r}$ arising from this Monte-Carlo based sampling is shown in Supplementary Fig.\,\ref{fig:fe_sim}b.

\subsection{Far-field collection model}
The emission from each dipole was calculated using the frequency-domain scattering solver of JCMSuite \cite{jcmwave}. Because the NV centers emit independently, their collected optical powers were summed incoherently. No fixed phase relation between different emitters was assumed. The electromagnetic field was evaluated on a plane located more than 1.3$\mu$m above the top surface of the cavity and transformed into its angular spectrum. The model assumes independent emitters, a static distribution of NV positions and orientations, and identical intrinsic radiative properties for all NV centers. Spectral diffusion, NV-to-NV variations in quantum efficiency, and non-radiative decay are not included.

The collected power was obtained by integrating over transverse wavevectors satisfying
$$
k_{\parallel}\leq k_0\mathrm{NA},
$$
where $k_{\parallel}$ is the magnitude of the wavevector component parallel to the collection plane, $k_0$ is the free-space wavevector and $\mathrm{NA}=0.82$ is the numerical aperture of the objective.

The collected power was calculated separately for each independent dipole source and subsequently summed over the ensemble.
\begin{figure}
    \centering
    \includegraphics[]{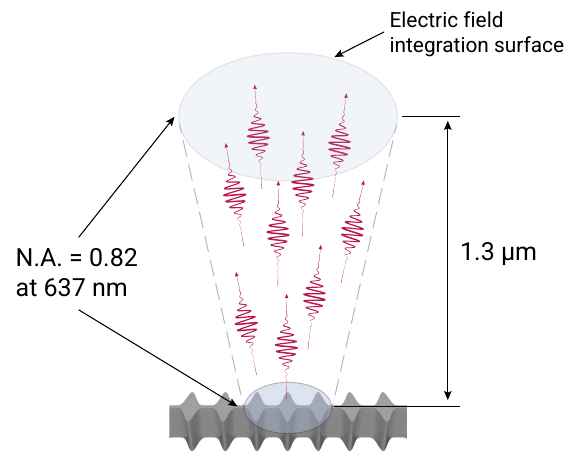}
    \caption{Visualization of the FEM simulation setup. Radiated intensity within the angular range defined by the numerical aperture of the objective is collected and contributes to the far-field signal.}
    \label{fig:farfield}
\end{figure}

\section{Supplementary Note 6: Gas tuning and multi-device alignment}

\subsection{Nitrogen delivery and condensation}
Ultra-pure nitrogen gas with a specified purity of at least 99.999\% was introduced into the cryostat through a fine-dosing valve (F3VCR-454, nenion leakvalves) and a capillary terminating close to the sample. At the sample temperature of approximately 5\,K, nitrogen condensed onto the device surface. The resulting dielectric loading increased the refractive index of the cavity environment and red-shifted the cavity resonance.

\subsection{Local laser-induced desorption}
Spatially selective blue-tuning was achieved by locally heating the condensed nitrogen layer with the focused 520\,nm laser. A power of approximately 2\,mW at the sample was applied in pulses of 50--500\,ms. After each pulse, the cavity spectrum was recorded and the pulse duration was adjusted as the resonance approached the target wavelength.

\subsection{Tuning range and cavity quality factor}
Individual nitrogen-deposition steps produced a maximum observed red-shift exceeding 25\,nm (Supplementary Fig.\,\ref{fig:exp_schematic}b). The cavity resonance and quality factor were extracted after each tuning step using the procedure described in Supplementary Note\,4. The measured $Q$ factors remained within $\pm\,500$ over the accessible tuning range (Supplementary Fig.\,\ref{fig:cavity_q}).

%% file: main.bbl
\begin{thebibliography}{36}%
\makeatletter
\providecommand \@ifxundefined [1]{%
 \@ifx{#1\undefined}
}%
\providecommand \@ifnum [1]{%
 \ifnum #1\expandafter \@firstoftwo
 \else \expandafter \@secondoftwo
 \fi
}%
\providecommand \@ifx [1]{%
 \ifx #1\expandafter \@firstoftwo
 \else \expandafter \@secondoftwo
 \fi
}%
\providecommand \natexlab [1]{#1}%
\providecommand \enquote  [1]{``#1''}%
\providecommand \bibnamefont  [1]{#1}%
\providecommand \bibfnamefont [1]{#1}%
\providecommand \citenamefont [1]{#1}%
\providecommand \href@noop [0]{\@secondoftwo}%
\providecommand \href [0]{\begingroup \@sanitize@url \@href}%
\providecommand \@href[1]{\@@startlink{#1}\@@href}%
\providecommand \@@href[1]{\endgroup#1\@@endlink}%
\providecommand \@sanitize@url [0]{\catcode `\\12\catcode `\$12\catcode
  `\&12\catcode `\#12\catcode `\^12\catcode `\_12\catcode `\%12\relax}%
\providecommand \@@startlink[1]{}%
\providecommand \@@endlink[0]{}%
\providecommand \url  [0]{\begingroup\@sanitize@url \@url }%
\providecommand \@url [1]{\endgroup\@href {#1}{\urlprefix }}%
\providecommand \urlprefix  [0]{URL }%
\providecommand \Eprint [0]{\href }%
\providecommand \doibase [0]{https://doi.org/}%
\providecommand \selectlanguage [0]{\@gobble}%
\providecommand \bibinfo  [0]{\@secondoftwo}%
\providecommand \bibfield  [0]{\@secondoftwo}%
\providecommand \translation [1]{[#1]}%
\providecommand \BibitemOpen [0]{}%
\providecommand \bibitemStop [0]{}%
\providecommand \bibitemNoStop [0]{.\EOS\space}%
\providecommand \EOS [0]{\spacefactor3000\relax}%
\providecommand \BibitemShut  [1]{\csname bibitem#1\endcsname}%
\let\auto@bib@innerbib\@empty
\bibitem [{\citenamefont {Oh}\ \emph {et~al.}(2024)\citenamefont {Oh},
  \citenamefont {Gregoire}, \citenamefont {Black}, \citenamefont {Hughes},
  \citenamefont {Kunz}, \citenamefont {Larsen}, \citenamefont {Lautier-Gaud},
  \citenamefont {Lee}, \citenamefont {Schwindt}, \citenamefont {Mouradian},
  \citenamefont {Narducci},\ and\ \citenamefont
  {Sackett}}]{oh_perspective_2024}%
  \BibitemOpen
  \bibfield  {author} {\bibinfo {author} {\bibfnamefont {E.}~\bibnamefont
  {Oh}}, \bibinfo {author} {\bibfnamefont {M.~D.}\ \bibnamefont {Gregoire}},
  \bibinfo {author} {\bibfnamefont {A.~T.}\ \bibnamefont {Black}}, \bibinfo
  {author} {\bibfnamefont {K.~J.}\ \bibnamefont {Hughes}}, \bibinfo {author}
  {\bibfnamefont {P.~D.}\ \bibnamefont {Kunz}}, \bibinfo {author}
  {\bibfnamefont {M.}~\bibnamefont {Larsen}}, \bibinfo {author} {\bibfnamefont
  {J.}~\bibnamefont {Lautier-Gaud}}, \bibinfo {author} {\bibfnamefont
  {J.}~\bibnamefont {Lee}}, \bibinfo {author} {\bibfnamefont {P.~D.~D.}\
  \bibnamefont {Schwindt}}, \bibinfo {author} {\bibfnamefont {S.~L.}\
  \bibnamefont {Mouradian}}, \bibinfo {author} {\bibfnamefont {F.~A.}\
  \bibnamefont {Narducci}},\ and\ \bibinfo {author} {\bibfnamefont {C.~A.}\
  \bibnamefont {Sackett}},\ }\href {https://doi.org/10.48550/ARXIV.2407.00689}
  {\bibinfo {title} {A {Perspective} on {Quantum} {Sensors} from {Basic}
  {Research} to {Commercial} {Applications}}} (\bibinfo {year} {2024}),\
  \Eprint {https://arxiv.org/abs/2407.00689} {arXiv:2407.00689 [quant-ph]}
  \BibitemShut {NoStop}%
\bibitem [{\citenamefont {Aquina}\ \emph {et~al.}(2025)\citenamefont {Aquina},
  \citenamefont {Cimoli}, \citenamefont {Das}, \citenamefont {Hövelmanns},
  \citenamefont {Weber}, \citenamefont {Okonkwo}, \citenamefont {Rommel},
  \citenamefont {Škorić}, \citenamefont {Tafur~Monroy},\ and\ \citenamefont
  {Verschoor}}]{aquina_critical_2025}%
  \BibitemOpen
  \bibfield  {author} {\bibinfo {author} {\bibfnamefont {N.}~\bibnamefont
  {Aquina}}, \bibinfo {author} {\bibfnamefont {B.}~\bibnamefont {Cimoli}},
  \bibinfo {author} {\bibfnamefont {S.}~\bibnamefont {Das}}, \bibinfo {author}
  {\bibfnamefont {K.}~\bibnamefont {Hövelmanns}}, \bibinfo {author}
  {\bibfnamefont {F.~J.}\ \bibnamefont {Weber}}, \bibinfo {author}
  {\bibfnamefont {C.}~\bibnamefont {Okonkwo}}, \bibinfo {author} {\bibfnamefont
  {S.}~\bibnamefont {Rommel}}, \bibinfo {author} {\bibfnamefont
  {B.}~\bibnamefont {Škorić}}, \bibinfo {author} {\bibfnamefont
  {I.}~\bibnamefont {Tafur~Monroy}},\ and\ \bibinfo {author} {\bibfnamefont
  {S.}~\bibnamefont {Verschoor}},\ }\href
  {https://doi.org/10.1140/epjqt/s40507-025-00350-5} {\bibfield  {journal}
  {\bibinfo  {journal} {EPJ Quantum Technology}\ }\textbf {\bibinfo {volume}
  {12}},\ \bibinfo {pages} {51} (\bibinfo {year} {2025})}\BibitemShut {NoStop}%
\bibitem [{\citenamefont {Seskir}\ \emph {et~al.}(2022)\citenamefont {Seskir},
  \citenamefont {Korkmaz},\ and\ \citenamefont
  {Aydinoglu}}]{seskir_landscape_2022}%
  \BibitemOpen
  \bibfield  {author} {\bibinfo {author} {\bibfnamefont {Z.~C.}\ \bibnamefont
  {Seskir}}, \bibinfo {author} {\bibfnamefont {R.}~\bibnamefont {Korkmaz}},\
  and\ \bibinfo {author} {\bibfnamefont {A.~U.}\ \bibnamefont {Aydinoglu}},\
  }\href {https://doi.org/10.1140/epjqt/s40507-022-00146-x} {\bibfield
  {journal} {\bibinfo  {journal} {EPJ Quantum Technology}\ }\textbf {\bibinfo
  {volume} {9}},\ \bibinfo {pages} {27} (\bibinfo {year} {2022})}\BibitemShut
  {NoStop}%
\bibitem [{\citenamefont {Bakr}\ \emph {et~al.}(2025)\citenamefont {Bakr},
  \citenamefont {Fasciati}, \citenamefont {Cao}, \citenamefont {Campanaro},
  \citenamefont {Wills}, \citenamefont {Alghadeer}, \citenamefont {Piscitelli},
  \citenamefont {Shteynas}, \citenamefont {Chidambaram},\ and\ \citenamefont
  {Leek}}]{bakr_multiplexed_2025}%
  \BibitemOpen
  \bibfield  {author} {\bibinfo {author} {\bibfnamefont {M.}~\bibnamefont
  {Bakr}}, \bibinfo {author} {\bibfnamefont {S.~D.}\ \bibnamefont {Fasciati}},
  \bibinfo {author} {\bibfnamefont {S.}~\bibnamefont {Cao}}, \bibinfo {author}
  {\bibfnamefont {G.}~\bibnamefont {Campanaro}}, \bibinfo {author}
  {\bibfnamefont {J.}~\bibnamefont {Wills}}, \bibinfo {author} {\bibfnamefont
  {M.}~\bibnamefont {Alghadeer}}, \bibinfo {author} {\bibfnamefont
  {M.}~\bibnamefont {Piscitelli}}, \bibinfo {author} {\bibfnamefont
  {B.}~\bibnamefont {Shteynas}}, \bibinfo {author} {\bibfnamefont
  {V.}~\bibnamefont {Chidambaram}},\ and\ \bibinfo {author} {\bibfnamefont
  {P.~J.}\ \bibnamefont {Leek}},\ }\href
  {https://doi.org/10.1103/PhysRevApplied.23.054089} {\bibfield  {journal}
  {\bibinfo  {journal} {Physical Review Applied}\ }\textbf {\bibinfo {volume}
  {23}},\ \bibinfo {pages} {054089} (\bibinfo {year} {2025})}\BibitemShut
  {NoStop}%
\bibitem [{\citenamefont {Atatüre}\ \emph {et~al.}(2018)\citenamefont
  {Atatüre}, \citenamefont {Englund}, \citenamefont {Vamivakas}, \citenamefont
  {Lee},\ and\ \citenamefont {Wrachtrup}}]{atature_material_2018}%
  \BibitemOpen
  \bibfield  {author} {\bibinfo {author} {\bibfnamefont {M.}~\bibnamefont
  {Atatüre}}, \bibinfo {author} {\bibfnamefont {D.}~\bibnamefont {Englund}},
  \bibinfo {author} {\bibfnamefont {N.}~\bibnamefont {Vamivakas}}, \bibinfo
  {author} {\bibfnamefont {S.-Y.}\ \bibnamefont {Lee}},\ and\ \bibinfo {author}
  {\bibfnamefont {J.}~\bibnamefont {Wrachtrup}},\ }\href
  {https://doi.org/10.1038/s41578-018-0008-9} {\bibfield  {journal} {\bibinfo
  {journal} {Nature Reviews Materials}\ }\textbf {\bibinfo {volume} {3}},\
  \bibinfo {pages} {38} (\bibinfo {year} {2018})}\BibitemShut {NoStop}%
\bibitem [{\citenamefont {Wehner}\ \emph {et~al.}(2018)\citenamefont {Wehner},
  \citenamefont {Elkouss},\ and\ \citenamefont {Hanson}}]{wehner_quantum_2018}%
  \BibitemOpen
  \bibfield  {author} {\bibinfo {author} {\bibfnamefont {S.}~\bibnamefont
  {Wehner}}, \bibinfo {author} {\bibfnamefont {D.}~\bibnamefont {Elkouss}},\
  and\ \bibinfo {author} {\bibfnamefont {R.}~\bibnamefont {Hanson}},\ }\href
  {https://doi.org/10.1126/science.aam9288} {\bibfield  {journal} {\bibinfo
  {journal} {Science}\ }\textbf {\bibinfo {volume} {362}},\ \bibinfo {pages}
  {eaam9288} (\bibinfo {year} {2018})}\BibitemShut {NoStop}%
\bibitem [{\citenamefont {Nickerson}\ \emph {et~al.}(2014)\citenamefont
  {Nickerson}, \citenamefont {Fitzsimons},\ and\ \citenamefont
  {Benjamin}}]{nickerson_freely_2014}%
  \BibitemOpen
  \bibfield  {author} {\bibinfo {author} {\bibfnamefont {N.~H.}\ \bibnamefont
  {Nickerson}}, \bibinfo {author} {\bibfnamefont {J.~F.}\ \bibnamefont
  {Fitzsimons}},\ and\ \bibinfo {author} {\bibfnamefont {S.~C.}\ \bibnamefont
  {Benjamin}},\ }\href {https://doi.org/10.1103/PhysRevX.4.041041} {\bibfield
  {journal} {\bibinfo  {journal} {Physical Review X}\ }\textbf {\bibinfo
  {volume} {4}},\ \bibinfo {pages} {041041} (\bibinfo {year}
  {2014})}\BibitemShut {NoStop}%
\bibitem [{\citenamefont {Basso~Basset}\ \emph {et~al.}(2021)\citenamefont
  {Basso~Basset}, \citenamefont {Valeri}, \citenamefont {Roccia}, \citenamefont
  {Muredda}, \citenamefont {Poderini}, \citenamefont {Neuwirth}, \citenamefont
  {Spagnolo}, \citenamefont {Rota}, \citenamefont {Carvacho}, \citenamefont
  {Sciarrino},\ and\ \citenamefont {Trotta}}]{basso_basset_quantum_2021}%
  \BibitemOpen
  \bibfield  {author} {\bibinfo {author} {\bibfnamefont {F.}~\bibnamefont
  {Basso~Basset}}, \bibinfo {author} {\bibfnamefont {M.}~\bibnamefont
  {Valeri}}, \bibinfo {author} {\bibfnamefont {E.}~\bibnamefont {Roccia}},
  \bibinfo {author} {\bibfnamefont {V.}~\bibnamefont {Muredda}}, \bibinfo
  {author} {\bibfnamefont {D.}~\bibnamefont {Poderini}}, \bibinfo {author}
  {\bibfnamefont {J.}~\bibnamefont {Neuwirth}}, \bibinfo {author}
  {\bibfnamefont {N.}~\bibnamefont {Spagnolo}}, \bibinfo {author}
  {\bibfnamefont {M.~B.}\ \bibnamefont {Rota}}, \bibinfo {author}
  {\bibfnamefont {G.}~\bibnamefont {Carvacho}}, \bibinfo {author}
  {\bibfnamefont {F.}~\bibnamefont {Sciarrino}},\ and\ \bibinfo {author}
  {\bibfnamefont {R.}~\bibnamefont {Trotta}},\ }\href
  {https://doi.org/10.1126/sciadv.abe6379} {\bibfield  {journal} {\bibinfo
  {journal} {Science Advances}\ }\textbf {\bibinfo {volume} {7}},\ \bibinfo
  {pages} {eabe6379} (\bibinfo {year} {2021})}\BibitemShut {NoStop}%
\bibitem [{\citenamefont {Ekert}\ and\ \citenamefont
  {Renner}(2014)}]{ekert_ultimate_2014}%
  \BibitemOpen
  \bibfield  {author} {\bibinfo {author} {\bibfnamefont {A.}~\bibnamefont
  {Ekert}}\ and\ \bibinfo {author} {\bibfnamefont {R.}~\bibnamefont {Renner}},\
  }\href {https://doi.org/10.1038/nature13132} {\bibfield  {journal} {\bibinfo
  {journal} {Nature}\ }\textbf {\bibinfo {volume} {507}},\ \bibinfo {pages}
  {443} (\bibinfo {year} {2014})}\BibitemShut {NoStop}%
\bibitem [{\citenamefont {Stas}\ \emph {et~al.}(2026)\citenamefont {Stas},
  \citenamefont {Wei}, \citenamefont {Sirotin}, \citenamefont {Huan},
  \citenamefont {Yazlar}, \citenamefont {Abdo~Arias}, \citenamefont {Knyazev},
  \citenamefont {Baranes}, \citenamefont {Machielse}, \citenamefont {Grandi},
  \citenamefont {Riedel}, \citenamefont {Borregaard}, \citenamefont {Park},
  \citenamefont {Lončar}, \citenamefont {Suleymanzade},\ and\ \citenamefont
  {Lukin}}]{stas_entanglement-assisted_2026}%
  \BibitemOpen
  \bibfield  {author} {\bibinfo {author} {\bibfnamefont {P.-J.}\ \bibnamefont
  {Stas}}, \bibinfo {author} {\bibfnamefont {Y.-C.}\ \bibnamefont {Wei}},
  \bibinfo {author} {\bibfnamefont {M.}~\bibnamefont {Sirotin}}, \bibinfo
  {author} {\bibfnamefont {Y.~Q.}\ \bibnamefont {Huan}}, \bibinfo {author}
  {\bibfnamefont {U.}~\bibnamefont {Yazlar}}, \bibinfo {author} {\bibfnamefont
  {F.}~\bibnamefont {Abdo~Arias}}, \bibinfo {author} {\bibfnamefont
  {E.}~\bibnamefont {Knyazev}}, \bibinfo {author} {\bibfnamefont
  {G.}~\bibnamefont {Baranes}}, \bibinfo {author} {\bibfnamefont
  {B.}~\bibnamefont {Machielse}}, \bibinfo {author} {\bibfnamefont
  {S.}~\bibnamefont {Grandi}}, \bibinfo {author} {\bibfnamefont
  {D.}~\bibnamefont {Riedel}}, \bibinfo {author} {\bibfnamefont
  {J.}~\bibnamefont {Borregaard}}, \bibinfo {author} {\bibfnamefont
  {H.}~\bibnamefont {Park}}, \bibinfo {author} {\bibfnamefont {M.}~\bibnamefont
  {Lončar}}, \bibinfo {author} {\bibfnamefont {A.}~\bibnamefont
  {Suleymanzade}},\ and\ \bibinfo {author} {\bibfnamefont {M.~D.}\ \bibnamefont
  {Lukin}},\ }\href {https://doi.org/10.1038/s41586-026-10171-w} {\bibfield
  {journal} {\bibinfo  {journal} {Nature}\ }\textbf {\bibinfo {volume} {651}},\
  \bibinfo {pages} {326} (\bibinfo {year} {2026})}\BibitemShut {NoStop}%
\bibitem [{\citenamefont {Balasubramanian}\ \emph {et~al.}(2009)\citenamefont
  {Balasubramanian}, \citenamefont {Neumann}, \citenamefont {Twitchen},
  \citenamefont {Markham}, \citenamefont {Kolesov}, \citenamefont {Mizuochi},
  \citenamefont {Isoya}, \citenamefont {Achard}, \citenamefont {Beck},
  \citenamefont {Tissler}, \citenamefont {Jacques}, \citenamefont {Hemmer},
  \citenamefont {Jelezko},\ and\ \citenamefont
  {Wrachtrup}}]{balasubramanian_ultralong_2009}%
  \BibitemOpen
  \bibfield  {author} {\bibinfo {author} {\bibfnamefont {G.}~\bibnamefont
  {Balasubramanian}}, \bibinfo {author} {\bibfnamefont {P.}~\bibnamefont
  {Neumann}}, \bibinfo {author} {\bibfnamefont {D.}~\bibnamefont {Twitchen}},
  \bibinfo {author} {\bibfnamefont {M.}~\bibnamefont {Markham}}, \bibinfo
  {author} {\bibfnamefont {R.}~\bibnamefont {Kolesov}}, \bibinfo {author}
  {\bibfnamefont {N.}~\bibnamefont {Mizuochi}}, \bibinfo {author}
  {\bibfnamefont {J.}~\bibnamefont {Isoya}}, \bibinfo {author} {\bibfnamefont
  {J.}~\bibnamefont {Achard}}, \bibinfo {author} {\bibfnamefont
  {J.}~\bibnamefont {Beck}}, \bibinfo {author} {\bibfnamefont {J.}~\bibnamefont
  {Tissler}}, \bibinfo {author} {\bibfnamefont {V.}~\bibnamefont {Jacques}},
  \bibinfo {author} {\bibfnamefont {P.~R.}\ \bibnamefont {Hemmer}}, \bibinfo
  {author} {\bibfnamefont {F.}~\bibnamefont {Jelezko}},\ and\ \bibinfo {author}
  {\bibfnamefont {J.}~\bibnamefont {Wrachtrup}},\ }\href
  {https://doi.org/10.1038/nmat2420} {\bibfield  {journal} {\bibinfo  {journal}
  {Nature Materials}\ }\textbf {\bibinfo {volume} {8}},\ \bibinfo {pages} {383}
  (\bibinfo {year} {2009})}\BibitemShut {NoStop}%
\bibitem [{\citenamefont {Maurer}\ \emph {et~al.}(2012)\citenamefont {Maurer},
  \citenamefont {Kucsko}, \citenamefont {Latta}, \citenamefont {Jiang},
  \citenamefont {Yao}, \citenamefont {Bennett}, \citenamefont {Pastawski},
  \citenamefont {Hunger}, \citenamefont {Chisholm}, \citenamefont {Markham},
  \citenamefont {Twitchen}, \citenamefont {Cirac},\ and\ \citenamefont
  {Lukin}}]{maurer_room-temperature_2012}%
  \BibitemOpen
  \bibfield  {author} {\bibinfo {author} {\bibfnamefont {P.~C.}\ \bibnamefont
  {Maurer}}, \bibinfo {author} {\bibfnamefont {G.}~\bibnamefont {Kucsko}},
  \bibinfo {author} {\bibfnamefont {C.}~\bibnamefont {Latta}}, \bibinfo
  {author} {\bibfnamefont {L.}~\bibnamefont {Jiang}}, \bibinfo {author}
  {\bibfnamefont {N.~Y.}\ \bibnamefont {Yao}}, \bibinfo {author} {\bibfnamefont
  {S.~D.}\ \bibnamefont {Bennett}}, \bibinfo {author} {\bibfnamefont
  {F.}~\bibnamefont {Pastawski}}, \bibinfo {author} {\bibfnamefont
  {D.}~\bibnamefont {Hunger}}, \bibinfo {author} {\bibfnamefont
  {N.}~\bibnamefont {Chisholm}}, \bibinfo {author} {\bibfnamefont
  {M.}~\bibnamefont {Markham}}, \bibinfo {author} {\bibfnamefont {D.~J.}\
  \bibnamefont {Twitchen}}, \bibinfo {author} {\bibfnamefont {J.~I.}\
  \bibnamefont {Cirac}},\ and\ \bibinfo {author} {\bibfnamefont {M.~D.}\
  \bibnamefont {Lukin}},\ }\href {https://doi.org/10.1126/science.1220513}
  {\bibfield  {journal} {\bibinfo  {journal} {Science}\ }\textbf {\bibinfo
  {volume} {336}},\ \bibinfo {pages} {1283} (\bibinfo {year}
  {2012})}\BibitemShut {NoStop}%
\bibitem [{\citenamefont {Abobeih}\ \emph {et~al.}(2018)\citenamefont
  {Abobeih}, \citenamefont {Cramer}, \citenamefont {Bakker}, \citenamefont
  {Kalb}, \citenamefont {Markham}, \citenamefont {Twitchen},\ and\
  \citenamefont {Taminiau}}]{abobeih_one-second_2018}%
  \BibitemOpen
  \bibfield  {author} {\bibinfo {author} {\bibfnamefont {M.~H.}\ \bibnamefont
  {Abobeih}}, \bibinfo {author} {\bibfnamefont {J.}~\bibnamefont {Cramer}},
  \bibinfo {author} {\bibfnamefont {M.~A.}\ \bibnamefont {Bakker}}, \bibinfo
  {author} {\bibfnamefont {N.}~\bibnamefont {Kalb}}, \bibinfo {author}
  {\bibfnamefont {M.}~\bibnamefont {Markham}}, \bibinfo {author} {\bibfnamefont
  {D.~J.}\ \bibnamefont {Twitchen}},\ and\ \bibinfo {author} {\bibfnamefont
  {T.~H.}\ \bibnamefont {Taminiau}},\ }\href
  {https://doi.org/10.1038/s41467-018-04916-z} {\bibfield  {journal} {\bibinfo
  {journal} {Nature Communications}\ }\textbf {\bibinfo {volume} {9}},\
  \bibinfo {pages} {2552} (\bibinfo {year} {2018})}\BibitemShut {NoStop}%
\bibitem [{\citenamefont {Bradley}\ \emph {et~al.}(2019)\citenamefont
  {Bradley}, \citenamefont {Randall}, \citenamefont {Abobeih}, \citenamefont
  {Berrevoets}, \citenamefont {Degen}, \citenamefont {Bakker}, \citenamefont
  {Markham}, \citenamefont {Twitchen},\ and\ \citenamefont
  {Taminiau}}]{bradley_ten-qubit_2019}%
  \BibitemOpen
  \bibfield  {author} {\bibinfo {author} {\bibfnamefont {C.}~\bibnamefont
  {Bradley}}, \bibinfo {author} {\bibfnamefont {J.}~\bibnamefont {Randall}},
  \bibinfo {author} {\bibfnamefont {M.}~\bibnamefont {Abobeih}}, \bibinfo
  {author} {\bibfnamefont {R.}~\bibnamefont {Berrevoets}}, \bibinfo {author}
  {\bibfnamefont {M.}~\bibnamefont {Degen}}, \bibinfo {author} {\bibfnamefont
  {M.}~\bibnamefont {Bakker}}, \bibinfo {author} {\bibfnamefont
  {M.}~\bibnamefont {Markham}}, \bibinfo {author} {\bibfnamefont
  {D.}~\bibnamefont {Twitchen}},\ and\ \bibinfo {author} {\bibfnamefont
  {T.}~\bibnamefont {Taminiau}},\ }\href
  {https://doi.org/10.1103/PhysRevX.9.031045} {\bibfield  {journal} {\bibinfo
  {journal} {Physical Review X}\ }\textbf {\bibinfo {volume} {9}},\ \bibinfo
  {pages} {031045} (\bibinfo {year} {2019})}\BibitemShut {NoStop}%
\bibitem [{\citenamefont {Yamamoto}\ \emph {et~al.}(2026)\citenamefont
  {Yamamoto}, \citenamefont {van Ommen}, \citenamefont {Schymik}, \citenamefont
  {de~Zoeten}, \citenamefont {Onoda}, \citenamefont {Saiki}, \citenamefont
  {Ohshima}, \citenamefont {Arjmandi-Tash}, \citenamefont {Vollmer},\ and\
  \citenamefont {Taminiau}}]{yamamoto_tensecond_2026}%
  \BibitemOpen
  \bibfield  {author} {\bibinfo {author} {\bibfnamefont {T.}~\bibnamefont
  {Yamamoto}}, \bibinfo {author} {\bibfnamefont {H.~B.}\ \bibnamefont {van
  Ommen}}, \bibinfo {author} {\bibfnamefont {K.-N.}\ \bibnamefont {Schymik}},
  \bibinfo {author} {\bibfnamefont {B.}~\bibnamefont {de~Zoeten}}, \bibinfo
  {author} {\bibfnamefont {S.}~\bibnamefont {Onoda}}, \bibinfo {author}
  {\bibfnamefont {S.}~\bibnamefont {Saiki}}, \bibinfo {author} {\bibfnamefont
  {T.}~\bibnamefont {Ohshima}}, \bibinfo {author} {\bibfnamefont
  {H.}~\bibnamefont {Arjmandi-Tash}}, \bibinfo {author} {\bibfnamefont
  {R.}~\bibnamefont {Vollmer}},\ and\ \bibinfo {author} {\bibfnamefont {T.~H.}\
  \bibnamefont {Taminiau}},\ }\href {https://arxiv.org/abs/2604.07439}
  {\bibinfo {title} {Ten-second electron-spin coherence in isotopically
  engineered diamond}} (\bibinfo {year} {2026}),\ \Eprint
  {https://arxiv.org/abs/2604.07439} {arXiv:2604.07439 [quant-ph]} \BibitemShut
  {NoStop}%
\bibitem [{\citenamefont {Pompili}\ \emph {et~al.}(2021)\citenamefont
  {Pompili}, \citenamefont {Hermans}, \citenamefont {Baier}, \citenamefont
  {Beukers}, \citenamefont {Humphreys}, \citenamefont {Schouten}, \citenamefont
  {Vermeulen}, \citenamefont {Tiggelman}, \citenamefont {Dos Santos~Martins},
  \citenamefont {Dirkse}, \citenamefont {Wehner},\ and\ \citenamefont
  {Hanson}}]{pompili_realization_2021}%
  \BibitemOpen
  \bibfield  {author} {\bibinfo {author} {\bibfnamefont {M.}~\bibnamefont
  {Pompili}}, \bibinfo {author} {\bibfnamefont {S.~L.~N.}\ \bibnamefont
  {Hermans}}, \bibinfo {author} {\bibfnamefont {S.}~\bibnamefont {Baier}},
  \bibinfo {author} {\bibfnamefont {H.~K.~C.}\ \bibnamefont {Beukers}},
  \bibinfo {author} {\bibfnamefont {P.~C.}\ \bibnamefont {Humphreys}}, \bibinfo
  {author} {\bibfnamefont {R.~N.}\ \bibnamefont {Schouten}}, \bibinfo {author}
  {\bibfnamefont {R.~F.~L.}\ \bibnamefont {Vermeulen}}, \bibinfo {author}
  {\bibfnamefont {M.~J.}\ \bibnamefont {Tiggelman}}, \bibinfo {author}
  {\bibfnamefont {L.}~\bibnamefont {Dos Santos~Martins}}, \bibinfo {author}
  {\bibfnamefont {B.}~\bibnamefont {Dirkse}}, \bibinfo {author} {\bibfnamefont
  {S.}~\bibnamefont {Wehner}},\ and\ \bibinfo {author} {\bibfnamefont
  {R.}~\bibnamefont {Hanson}},\ }\href
  {https://doi.org/10.1126/science.abg1919} {\bibfield  {journal} {\bibinfo
  {journal} {Science}\ }\textbf {\bibinfo {volume} {372}},\ \bibinfo {pages}
  {259} (\bibinfo {year} {2021})}\BibitemShut {NoStop}%
\bibitem [{\citenamefont {Hausmann}\ \emph {et~al.}(2013)\citenamefont
  {Hausmann}, \citenamefont {Shields}, \citenamefont {Quan}, \citenamefont
  {Chu}, \citenamefont {De~Leon}, \citenamefont {Evans}, \citenamefont {Burek},
  \citenamefont {Zibrov}, \citenamefont {Markham}, \citenamefont {Twitchen},
  \citenamefont {Park}, \citenamefont {Lukin},\ and\ \citenamefont
  {Loncǎr}}]{hausmann_coupling_2013}%
  \BibitemOpen
  \bibfield  {author} {\bibinfo {author} {\bibfnamefont {B.~J.~M.}\
  \bibnamefont {Hausmann}}, \bibinfo {author} {\bibfnamefont {B.~J.}\
  \bibnamefont {Shields}}, \bibinfo {author} {\bibfnamefont {Q.}~\bibnamefont
  {Quan}}, \bibinfo {author} {\bibfnamefont {Y.}~\bibnamefont {Chu}}, \bibinfo
  {author} {\bibfnamefont {N.~P.}\ \bibnamefont {De~Leon}}, \bibinfo {author}
  {\bibfnamefont {R.}~\bibnamefont {Evans}}, \bibinfo {author} {\bibfnamefont
  {M.~J.}\ \bibnamefont {Burek}}, \bibinfo {author} {\bibfnamefont {A.~S.}\
  \bibnamefont {Zibrov}}, \bibinfo {author} {\bibfnamefont {M.}~\bibnamefont
  {Markham}}, \bibinfo {author} {\bibfnamefont {D.~J.}\ \bibnamefont
  {Twitchen}}, \bibinfo {author} {\bibfnamefont {H.}~\bibnamefont {Park}},
  \bibinfo {author} {\bibfnamefont {M.~D.}\ \bibnamefont {Lukin}},\ and\
  \bibinfo {author} {\bibfnamefont {M.}~\bibnamefont {Loncǎr}},\ }\href
  {https://doi.org/10.1021/nl402174g} {\bibfield  {journal} {\bibinfo
  {journal} {Nano Letters}\ }\textbf {\bibinfo {volume} {13}},\ \bibinfo
  {pages} {5791} (\bibinfo {year} {2013})}\BibitemShut {NoStop}%
\bibitem [{\citenamefont {Mouradian}\ \emph {et~al.}(2017)\citenamefont
  {Mouradian}, \citenamefont {Wan}, \citenamefont {Schröder},\ and\
  \citenamefont {Englund}}]{mouradian_rectangular_2017}%
  \BibitemOpen
  \bibfield  {author} {\bibinfo {author} {\bibfnamefont {S.}~\bibnamefont
  {Mouradian}}, \bibinfo {author} {\bibfnamefont {N.~H.}\ \bibnamefont {Wan}},
  \bibinfo {author} {\bibfnamefont {T.}~\bibnamefont {Schröder}},\ and\
  \bibinfo {author} {\bibfnamefont {D.}~\bibnamefont {Englund}},\ }\href
  {https://doi.org/10.1063/1.4992118} {\bibfield  {journal} {\bibinfo
  {journal} {Applied Physics Letters}\ }\textbf {\bibinfo {volume} {111}},\
  \bibinfo {pages} {021103} (\bibinfo {year} {2017})}\BibitemShut {NoStop}%
\bibitem [{\citenamefont {Li}\ \emph {et~al.}(2015)\citenamefont {Li},
  \citenamefont {Schröder}, \citenamefont {Chen}, \citenamefont {Walsh},
  \citenamefont {Bayn}, \citenamefont {Goldstein}, \citenamefont {Gaathon},
  \citenamefont {Trusheim}, \citenamefont {Lu}, \citenamefont {Mower},
  \citenamefont {Cotlet}, \citenamefont {Markham}, \citenamefont {Twitchen},\
  and\ \citenamefont {Englund}}]{li_coherent_2015}%
  \BibitemOpen
  \bibfield  {author} {\bibinfo {author} {\bibfnamefont {L.}~\bibnamefont
  {Li}}, \bibinfo {author} {\bibfnamefont {T.}~\bibnamefont {Schröder}},
  \bibinfo {author} {\bibfnamefont {E.~H.}\ \bibnamefont {Chen}}, \bibinfo
  {author} {\bibfnamefont {M.}~\bibnamefont {Walsh}}, \bibinfo {author}
  {\bibfnamefont {I.}~\bibnamefont {Bayn}}, \bibinfo {author} {\bibfnamefont
  {J.}~\bibnamefont {Goldstein}}, \bibinfo {author} {\bibfnamefont
  {O.}~\bibnamefont {Gaathon}}, \bibinfo {author} {\bibfnamefont {M.~E.}\
  \bibnamefont {Trusheim}}, \bibinfo {author} {\bibfnamefont {M.}~\bibnamefont
  {Lu}}, \bibinfo {author} {\bibfnamefont {J.}~\bibnamefont {Mower}}, \bibinfo
  {author} {\bibfnamefont {M.}~\bibnamefont {Cotlet}}, \bibinfo {author}
  {\bibfnamefont {M.~L.}\ \bibnamefont {Markham}}, \bibinfo {author}
  {\bibfnamefont {D.~J.}\ \bibnamefont {Twitchen}},\ and\ \bibinfo {author}
  {\bibfnamefont {D.}~\bibnamefont {Englund}},\ }\href
  {https://doi.org/10.1038/ncomms7173} {\bibfield  {journal} {\bibinfo
  {journal} {Nature Communications}\ }\textbf {\bibinfo {volume} {6}},\
  \bibinfo {pages} {6173} (\bibinfo {year} {2015})}\BibitemShut {NoStop}%
\bibitem [{\citenamefont {Faraon}\ \emph {et~al.}(2012)\citenamefont {Faraon},
  \citenamefont {Santori}, \citenamefont {Huang}, \citenamefont {Acosta},\ and\
  \citenamefont {Beausoleil}}]{faraon_coupling_2012}%
  \BibitemOpen
  \bibfield  {author} {\bibinfo {author} {\bibfnamefont {A.}~\bibnamefont
  {Faraon}}, \bibinfo {author} {\bibfnamefont {C.}~\bibnamefont {Santori}},
  \bibinfo {author} {\bibfnamefont {Z.}~\bibnamefont {Huang}}, \bibinfo
  {author} {\bibfnamefont {V.~M.}\ \bibnamefont {Acosta}},\ and\ \bibinfo
  {author} {\bibfnamefont {R.~G.}\ \bibnamefont {Beausoleil}},\ }\href
  {https://doi.org/10.1103/PhysRevLett.109.033604} {\bibfield  {journal}
  {\bibinfo  {journal} {Physical Review Letters}\ }\textbf {\bibinfo {volume}
  {109}},\ \bibinfo {pages} {033604} (\bibinfo {year} {2012})}\BibitemShut
  {NoStop}%
\bibitem [{\citenamefont {Stolk}\ \emph {et~al.}(2024)\citenamefont {Stolk},
  \citenamefont {Van Der~Enden}, \citenamefont {Slater}, \citenamefont
  {Te~Raa-Derckx}, \citenamefont {Botma}, \citenamefont {Van~Rantwijk},
  \citenamefont {Biemond}, \citenamefont {Hagen}, \citenamefont {Herfst},
  \citenamefont {Koek}, \citenamefont {Meskers}, \citenamefont {Vollmer},
  \citenamefont {Van~Zwet}, \citenamefont {Markham}, \citenamefont {Edmonds},
  \citenamefont {Geus}, \citenamefont {Elsen}, \citenamefont {Jungbluth},
  \citenamefont {Haefner}, \citenamefont {Tresp}, \citenamefont {Stuhler},
  \citenamefont {Ritter},\ and\ \citenamefont
  {Hanson}}]{stolk_metropolitan-scale_2024}%
  \BibitemOpen
  \bibfield  {author} {\bibinfo {author} {\bibfnamefont {A.~J.}\ \bibnamefont
  {Stolk}}, \bibinfo {author} {\bibfnamefont {K.~L.}\ \bibnamefont {Van
  Der~Enden}}, \bibinfo {author} {\bibfnamefont {M.-C.}\ \bibnamefont
  {Slater}}, \bibinfo {author} {\bibfnamefont {I.}~\bibnamefont
  {Te~Raa-Derckx}}, \bibinfo {author} {\bibfnamefont {P.}~\bibnamefont
  {Botma}}, \bibinfo {author} {\bibfnamefont {J.}~\bibnamefont {Van~Rantwijk}},
  \bibinfo {author} {\bibfnamefont {J.~J.~B.}\ \bibnamefont {Biemond}},
  \bibinfo {author} {\bibfnamefont {R.~A.~J.}\ \bibnamefont {Hagen}}, \bibinfo
  {author} {\bibfnamefont {R.~W.}\ \bibnamefont {Herfst}}, \bibinfo {author}
  {\bibfnamefont {W.~D.}\ \bibnamefont {Koek}}, \bibinfo {author}
  {\bibfnamefont {A.~J.~H.}\ \bibnamefont {Meskers}}, \bibinfo {author}
  {\bibfnamefont {R.}~\bibnamefont {Vollmer}}, \bibinfo {author} {\bibfnamefont
  {E.~J.}\ \bibnamefont {Van~Zwet}}, \bibinfo {author} {\bibfnamefont
  {M.}~\bibnamefont {Markham}}, \bibinfo {author} {\bibfnamefont {A.~M.}\
  \bibnamefont {Edmonds}}, \bibinfo {author} {\bibfnamefont {J.~F.}\
  \bibnamefont {Geus}}, \bibinfo {author} {\bibfnamefont {F.}~\bibnamefont
  {Elsen}}, \bibinfo {author} {\bibfnamefont {B.}~\bibnamefont {Jungbluth}},
  \bibinfo {author} {\bibfnamefont {C.}~\bibnamefont {Haefner}}, \bibinfo
  {author} {\bibfnamefont {C.}~\bibnamefont {Tresp}}, \bibinfo {author}
  {\bibfnamefont {J.}~\bibnamefont {Stuhler}}, \bibinfo {author} {\bibfnamefont
  {S.}~\bibnamefont {Ritter}},\ and\ \bibinfo {author} {\bibfnamefont
  {R.}~\bibnamefont {Hanson}},\ }\href {https://doi.org/10.1126/sciadv.adp6442}
  {\bibfield  {journal} {\bibinfo  {journal} {Science Advances}\ }\textbf
  {\bibinfo {volume} {10}},\ \bibinfo {pages} {eadp6442} (\bibinfo {year}
  {2024})}\BibitemShut {NoStop}%
\bibitem [{\citenamefont {Bopp}\ \emph {et~al.}(2024)\citenamefont {Bopp},
  \citenamefont {Plock}, \citenamefont {Turan}, \citenamefont {Pieplow},
  \citenamefont {Burger},\ and\ \citenamefont {Schröder}}]{bopp_sawfish_2024}%
  \BibitemOpen
  \bibfield  {author} {\bibinfo {author} {\bibfnamefont {J.~M.}\ \bibnamefont
  {Bopp}}, \bibinfo {author} {\bibfnamefont {M.}~\bibnamefont {Plock}},
  \bibinfo {author} {\bibfnamefont {T.}~\bibnamefont {Turan}}, \bibinfo
  {author} {\bibfnamefont {G.}~\bibnamefont {Pieplow}}, \bibinfo {author}
  {\bibfnamefont {S.}~\bibnamefont {Burger}},\ and\ \bibinfo {author}
  {\bibfnamefont {T.}~\bibnamefont {Schröder}},\ }\bibfield  {journal}
  {\bibinfo  {journal} {Advanced Optical Materials}\ }\textbf {\bibinfo
  {volume} {12}},\ \href {https://doi.org/10.1002/adom.202301286}
  {10.1002/adom.202301286} (\bibinfo {year} {2024})\BibitemShut {NoStop}%
\bibitem [{\citenamefont {Orphal-Kobin}\ \emph {et~al.}(2023)\citenamefont
  {Orphal-Kobin}, \citenamefont {Unterguggenberger}, \citenamefont
  {Pregnolato}, \citenamefont {Kemf}, \citenamefont {Matalla}, \citenamefont
  {Unger}, \citenamefont {Ostermay}, \citenamefont {Pieplow},\ and\
  \citenamefont {Schröder}}]{orphal-kobin_optically_2023}%
  \BibitemOpen
  \bibfield  {author} {\bibinfo {author} {\bibfnamefont {L.}~\bibnamefont
  {Orphal-Kobin}}, \bibinfo {author} {\bibfnamefont {K.}~\bibnamefont
  {Unterguggenberger}}, \bibinfo {author} {\bibfnamefont {T.}~\bibnamefont
  {Pregnolato}}, \bibinfo {author} {\bibfnamefont {N.}~\bibnamefont {Kemf}},
  \bibinfo {author} {\bibfnamefont {M.}~\bibnamefont {Matalla}}, \bibinfo
  {author} {\bibfnamefont {R.-S.}\ \bibnamefont {Unger}}, \bibinfo {author}
  {\bibfnamefont {I.}~\bibnamefont {Ostermay}}, \bibinfo {author}
  {\bibfnamefont {G.}~\bibnamefont {Pieplow}},\ and\ \bibinfo {author}
  {\bibfnamefont {T.}~\bibnamefont {Schröder}},\ }\bibfield  {journal}
  {\bibinfo  {journal} {Physical Review X}\ }\textbf {\bibinfo {volume} {13}},\
  \href {https://doi.org/10.1103/physrevx.13.011042}
  {10.1103/physrevx.13.011042} (\bibinfo {year} {2023})\BibitemShut {NoStop}%
\bibitem [{\citenamefont {Pregnolato}\ \emph {et~al.}(2024)\citenamefont
  {Pregnolato}, \citenamefont {Stucki}, \citenamefont {Bopp}, \citenamefont
  {V.~D.~Hoeven}, \citenamefont {Gokhale}, \citenamefont {Krüger},\ and\
  \citenamefont {Schröder}}]{pregnolato_fabrication_2024}%
  \BibitemOpen
  \bibfield  {author} {\bibinfo {author} {\bibfnamefont {T.}~\bibnamefont
  {Pregnolato}}, \bibinfo {author} {\bibfnamefont {M.~E.}\ \bibnamefont
  {Stucki}}, \bibinfo {author} {\bibfnamefont {J.~M.}\ \bibnamefont {Bopp}},
  \bibinfo {author} {\bibfnamefont {M.~H.}\ \bibnamefont {V.~D.~Hoeven}},
  \bibinfo {author} {\bibfnamefont {A.}~\bibnamefont {Gokhale}}, \bibinfo
  {author} {\bibfnamefont {O.}~\bibnamefont {Krüger}},\ and\ \bibinfo {author}
  {\bibfnamefont {T.}~\bibnamefont {Schröder}},\ }\bibfield  {journal}
  {\bibinfo  {journal} {APL Photonics}\ }\textbf {\bibinfo {volume} {9}},\
  \href {https://doi.org/10.1063/5.0186509} {10.1063/5.0186509} (\bibinfo
  {year} {2024})\BibitemShut {NoStop}%
\bibitem [{E6_()}]{E6_website}%
  \BibitemOpen
  \href@noop {} {\bibinfo {title} {{E}lement {S}ix {T}echnologies {L}td.
  ({UK}), {S}{C} {P}late {C}{V}{D} 3.0x3.0mm, 0.25mm thick, {P}{L}}},\ \bibinfo
  {howpublished}
  {\url{https://e6cvd.com/eur/material/single-crystalline/sc-plate-cvd-3-0x3-0x0-25mm-100-pl.html}}\BibitemShut
  {NoStop}%
\bibitem [{\citenamefont {Bauch}\ \emph {et~al.}(2020)\citenamefont {Bauch},
  \citenamefont {Singh}, \citenamefont {Lee}, \citenamefont {Hart},
  \citenamefont {Schloss}, \citenamefont {Turner}, \citenamefont {Barry},
  \citenamefont {Pham}, \citenamefont {Bar-Gill}, \citenamefont {Yelin},\ and\
  \citenamefont {Walsworth}}]{bauch_degoherence_2020}%
  \BibitemOpen
  \bibfield  {author} {\bibinfo {author} {\bibfnamefont {E.}~\bibnamefont
  {Bauch}}, \bibinfo {author} {\bibfnamefont {S.}~\bibnamefont {Singh}},
  \bibinfo {author} {\bibfnamefont {J.}~\bibnamefont {Lee}}, \bibinfo {author}
  {\bibfnamefont {C.~A.}\ \bibnamefont {Hart}}, \bibinfo {author}
  {\bibfnamefont {J.~M.}\ \bibnamefont {Schloss}}, \bibinfo {author}
  {\bibfnamefont {M.~J.}\ \bibnamefont {Turner}}, \bibinfo {author}
  {\bibfnamefont {J.~F.}\ \bibnamefont {Barry}}, \bibinfo {author}
  {\bibfnamefont {L.~M.}\ \bibnamefont {Pham}}, \bibinfo {author}
  {\bibfnamefont {N.}~\bibnamefont {Bar-Gill}}, \bibinfo {author}
  {\bibfnamefont {S.~F.}\ \bibnamefont {Yelin}},\ and\ \bibinfo {author}
  {\bibfnamefont {R.~L.}\ \bibnamefont {Walsworth}},\ }\href
  {https://doi.org/10.1103/PhysRevB.102.134210} {\bibfield  {journal} {\bibinfo
   {journal} {Phys. Rev. B}\ }\textbf {\bibinfo {volume} {102}},\ \bibinfo
  {pages} {134210} (\bibinfo {year} {2020})}\BibitemShut {NoStop}%
\bibitem [{\citenamefont {Orphal‐Kobin}\ \emph {et~al.}(2025)\citenamefont
  {Orphal‐Kobin}, \citenamefont {Torun}, \citenamefont {Bopp}, \citenamefont
  {Pieplow},\ and\ \citenamefont {Schröder}}]{orphalkobin_coherent_2025}%
  \BibitemOpen
  \bibfield  {author} {\bibinfo {author} {\bibfnamefont {L.}~\bibnamefont
  {Orphal‐Kobin}}, \bibinfo {author} {\bibfnamefont {C.~G.}\ \bibnamefont
  {Torun}}, \bibinfo {author} {\bibfnamefont {J.~M.}\ \bibnamefont {Bopp}},
  \bibinfo {author} {\bibfnamefont {G.}~\bibnamefont {Pieplow}},\ and\ \bibinfo
  {author} {\bibfnamefont {T.}~\bibnamefont {Schröder}},\ }\href
  {https://doi.org/10.1002/qute.202300432} {\bibfield  {journal} {\bibinfo
  {journal} {Advanced Quantum Technologies}\ }\textbf {\bibinfo {volume} {8}},\
  \bibinfo {pages} {2300432} (\bibinfo {year} {2025})}\BibitemShut {NoStop}%
\bibitem [{\citenamefont {Orphal-Kobin}\ \emph {et~al.}(2025)\citenamefont
  {Orphal-Kobin}, \citenamefont {Pieplow}, \citenamefont {Gokhale},
  \citenamefont {Unterguggenberger},\ and\ \citenamefont
  {Schröder}}]{orphal-kobin_retrieving_2025}%
  \BibitemOpen
  \bibfield  {author} {\bibinfo {author} {\bibfnamefont {L.}~\bibnamefont
  {Orphal-Kobin}}, \bibinfo {author} {\bibfnamefont {G.}~\bibnamefont
  {Pieplow}}, \bibinfo {author} {\bibfnamefont {A.}~\bibnamefont {Gokhale}},
  \bibinfo {author} {\bibfnamefont {K.}~\bibnamefont {Unterguggenberger}},\
  and\ \bibinfo {author} {\bibfnamefont {T.}~\bibnamefont {Schröder}},\ }\href
  {https://doi.org/10.48550/ARXIV.2501.07951} {\bibinfo {title} {Retrieving
  {Lost} {Atomic} {Information}: {Monte} {Carlo}-based {Parameter}
  {Reconstruction} of an {Optical} {Quantum} {System}}} (\bibinfo {year}
  {2025}),\ \Eprint {https://arxiv.org/abs/2501.07951} {arXiv:2501.07951
  [quant-ph]} \BibitemShut {NoStop}%
\bibitem [{\citenamefont {Mosor}\ \emph {et~al.}(2005)\citenamefont {Mosor},
  \citenamefont {Hendrickson}, \citenamefont {Richards}, \citenamefont {Sweet},
  \citenamefont {Khitrova}, \citenamefont {Gibbs}, \citenamefont {Yoshie},
  \citenamefont {Scherer}, \citenamefont {Shchekin},\ and\ \citenamefont
  {Deppe}}]{mosor_scanning_2005}%
  \BibitemOpen
  \bibfield  {author} {\bibinfo {author} {\bibfnamefont {S.}~\bibnamefont
  {Mosor}}, \bibinfo {author} {\bibfnamefont {J.}~\bibnamefont {Hendrickson}},
  \bibinfo {author} {\bibfnamefont {B.~C.}\ \bibnamefont {Richards}}, \bibinfo
  {author} {\bibfnamefont {J.}~\bibnamefont {Sweet}}, \bibinfo {author}
  {\bibfnamefont {G.}~\bibnamefont {Khitrova}}, \bibinfo {author}
  {\bibfnamefont {H.~M.}\ \bibnamefont {Gibbs}}, \bibinfo {author}
  {\bibfnamefont {T.}~\bibnamefont {Yoshie}}, \bibinfo {author} {\bibfnamefont
  {A.}~\bibnamefont {Scherer}}, \bibinfo {author} {\bibfnamefont {O.~B.}\
  \bibnamefont {Shchekin}},\ and\ \bibinfo {author} {\bibfnamefont {D.~G.}\
  \bibnamefont {Deppe}},\ }\href {https://doi.org/10.1063/1.2076435} {\bibfield
   {journal} {\bibinfo  {journal} {Applied Physics Letters}\ }\textbf {\bibinfo
  {volume} {87}},\ \bibinfo {pages} {141105} (\bibinfo {year}
  {2005})}\BibitemShut {NoStop}%
\bibitem [{\citenamefont {Rugar}\ \emph {et~al.}(2021)\citenamefont {Rugar},
  \citenamefont {Aghaeimeibodi}, \citenamefont {Riedel}, \citenamefont {Dory},
  \citenamefont {Lu}, \citenamefont {McQuade}, \citenamefont {Shen},
  \citenamefont {Melosh},\ and\ \citenamefont
  {Vučković}}]{rugar_quantum_2021}%
  \BibitemOpen
  \bibfield  {author} {\bibinfo {author} {\bibfnamefont {A.~E.}\ \bibnamefont
  {Rugar}}, \bibinfo {author} {\bibfnamefont {S.}~\bibnamefont
  {Aghaeimeibodi}}, \bibinfo {author} {\bibfnamefont {D.}~\bibnamefont
  {Riedel}}, \bibinfo {author} {\bibfnamefont {C.}~\bibnamefont {Dory}},
  \bibinfo {author} {\bibfnamefont {H.}~\bibnamefont {Lu}}, \bibinfo {author}
  {\bibfnamefont {P.~J.}\ \bibnamefont {McQuade}}, \bibinfo {author}
  {\bibfnamefont {Z.-X.}\ \bibnamefont {Shen}}, \bibinfo {author}
  {\bibfnamefont {N.~A.}\ \bibnamefont {Melosh}},\ and\ \bibinfo {author}
  {\bibfnamefont {J.}~\bibnamefont {Vučković}},\ }\href
  {https://doi.org/10.1103/PhysRevX.11.031021} {\bibfield  {journal} {\bibinfo
  {journal} {Physical Review X}\ }\textbf {\bibinfo {volume} {11}},\ \bibinfo
  {pages} {031021} (\bibinfo {year} {2021})}\BibitemShut {NoStop}%
\bibitem [{jcm()}]{jcmwave}%
  \BibitemOpen
  \href@noop {} {\bibinfo {title} {{J}{C}{M}suite}},\ \bibinfo {howpublished}
  {\url{https://jcmwave.com/jcmsuite}},\ \bibinfo {note} {{J}CMwave GmbH,
  Bolivarallee 22, 14050 Berlin, Germany}\BibitemShut {NoStop}%
\bibitem [{\citenamefont {Debye}(1913)}]{debye_interferenz_1913}%
  \BibitemOpen
  \bibfield  {author} {\bibinfo {author} {\bibfnamefont {P.}~\bibnamefont
  {Debye}},\ }\href {https://doi.org/10.1002/andp.19133480105} {\bibfield
  {journal} {\bibinfo  {journal} {Annalen der Physik}\ }\textbf {\bibinfo
  {volume} {348}},\ \bibinfo {pages} {49} (\bibinfo {year} {1913})}\BibitemShut
  {NoStop}%
\bibitem [{\citenamefont {Pieplow}\ \emph {et~al.}(2025)\citenamefont
  {Pieplow}, \citenamefont {Torun}, \citenamefont {Gurr}, \citenamefont
  {Munns}, \citenamefont {Herrmann}, \citenamefont {Thies}, \citenamefont
  {Pregnolato},\ and\ \citenamefont {Schröder}}]{pieplow_quantum_2025}%
  \BibitemOpen
  \bibfield  {author} {\bibinfo {author} {\bibfnamefont {G.}~\bibnamefont
  {Pieplow}}, \bibinfo {author} {\bibfnamefont {C.~G.}\ \bibnamefont {Torun}},
  \bibinfo {author} {\bibfnamefont {C.}~\bibnamefont {Gurr}}, \bibinfo {author}
  {\bibfnamefont {J.~H.~D.}\ \bibnamefont {Munns}}, \bibinfo {author}
  {\bibfnamefont {F.~M.}\ \bibnamefont {Herrmann}}, \bibinfo {author}
  {\bibfnamefont {A.}~\bibnamefont {Thies}}, \bibinfo {author} {\bibfnamefont
  {T.}~\bibnamefont {Pregnolato}},\ and\ \bibinfo {author} {\bibfnamefont
  {T.}~\bibnamefont {Schröder}},\ }\href
  {https://doi.org/10.1038/s41467-025-61839-2} {\bibfield  {journal} {\bibinfo
  {journal} {Nature Communications}\ }\textbf {\bibinfo {volume} {16}},\
  \bibinfo {pages} {6435} (\bibinfo {year} {2025})}\BibitemShut {NoStop}%
\bibitem [{\citenamefont {Alkauskas}\ \emph {et~al.}(2025)\citenamefont
  {Alkauskas}, \citenamefont {Van~de Walle}, \citenamefont {Razinkovas},\ and\
  \citenamefont {Ulbricht}}]{alkauskas_charge-state_2025}%
  \BibitemOpen
  \bibfield  {author} {\bibinfo {author} {\bibfnamefont {A.}~\bibnamefont
  {Alkauskas}}, \bibinfo {author} {\bibfnamefont {C.~G.}\ \bibnamefont {Van~de
  Walle}}, \bibinfo {author} {\bibfnamefont {L.}~\bibnamefont {Razinkovas}},\
  and\ \bibinfo {author} {\bibfnamefont {R.}~\bibnamefont {Ulbricht}},\
  }\href@noop {} {\bibinfo {title} {Charge state equilibration of
  nitrogen-vacancy center ensembles in diamond: The role of electron
  tunneling}} (\bibinfo {year} {2025}),\ \Eprint
  {https://arxiv.org/abs/2512.00952} {arXiv:2512.00952 [cond-mat.mtrl-sci]}
  \BibitemShut {NoStop}%
\bibitem [{\citenamefont {Manson}\ \emph {et~al.}(2026)\citenamefont {Manson},
  \citenamefont {Hedges}, \citenamefont {Barson}, \citenamefont {Meriles},
  \citenamefont {Ulbricht},\ and\ \citenamefont {Doherty}}]{manson_spin_2026}%
  \BibitemOpen
  \bibfield  {author} {\bibinfo {author} {\bibfnamefont {N.~B.}\ \bibnamefont
  {Manson}}, \bibinfo {author} {\bibfnamefont {M.}~\bibnamefont {Hedges}},
  \bibinfo {author} {\bibfnamefont {M.~S.~J.}\ \bibnamefont {Barson}}, \bibinfo
  {author} {\bibfnamefont {C.~A.}\ \bibnamefont {Meriles}}, \bibinfo {author}
  {\bibfnamefont {R.}~\bibnamefont {Ulbricht}},\ and\ \bibinfo {author}
  {\bibfnamefont {M.~W.}\ \bibnamefont {Doherty}},\ }\href@noop {} {\bibinfo
  {title} {Spin-dependent charge-state conversion in {NV} ensembles mediated by
  electron tunneling}} (\bibinfo {year} {2026}),\ \Eprint
  {https://arxiv.org/abs/2604.13337} {arXiv:2604.13337 [cond-mat.mes-hall]}
  \BibitemShut {NoStop}%
\bibitem [{\citenamefont {Burek}\ \emph {et~al.}(2017)\citenamefont {Burek},
  \citenamefont {Meuwly}, \citenamefont {Evans}, \citenamefont {Bhaskar},
  \citenamefont {Sipahigil}, \citenamefont {Meesala}, \citenamefont
  {Machielse}, \citenamefont {Sukachev}, \citenamefont {Nguyen}, \citenamefont
  {Pacheco}, \citenamefont {Bielejec}, \citenamefont {Lukin},\ and\
  \citenamefont {Lon\ifmmode~\check{c}\else
  \v{c}\fi{}ar}}]{burek_fiber-coupled_2017}%
  \BibitemOpen
  \bibfield  {author} {\bibinfo {author} {\bibfnamefont {M.~J.}\ \bibnamefont
  {Burek}}, \bibinfo {author} {\bibfnamefont {C.}~\bibnamefont {Meuwly}},
  \bibinfo {author} {\bibfnamefont {R.~E.}\ \bibnamefont {Evans}}, \bibinfo
  {author} {\bibfnamefont {M.~K.}\ \bibnamefont {Bhaskar}}, \bibinfo {author}
  {\bibfnamefont {A.}~\bibnamefont {Sipahigil}}, \bibinfo {author}
  {\bibfnamefont {S.}~\bibnamefont {Meesala}}, \bibinfo {author} {\bibfnamefont
  {B.}~\bibnamefont {Machielse}}, \bibinfo {author} {\bibfnamefont {D.~D.}\
  \bibnamefont {Sukachev}}, \bibinfo {author} {\bibfnamefont {C.~T.}\
  \bibnamefont {Nguyen}}, \bibinfo {author} {\bibfnamefont {J.~L.}\
  \bibnamefont {Pacheco}}, \bibinfo {author} {\bibfnamefont {E.}~\bibnamefont
  {Bielejec}}, \bibinfo {author} {\bibfnamefont {M.~D.}\ \bibnamefont
  {Lukin}},\ and\ \bibinfo {author} {\bibfnamefont {M.}~\bibnamefont
  {Lon\ifmmode~\check{c}\else \v{c}\fi{}ar}},\ }\href
  {https://doi.org/10.1103/PhysRevApplied.8.024026} {\bibfield  {journal}
  {\bibinfo  {journal} {Phys. Rev. Appl.}\ }\textbf {\bibinfo {volume} {8}},\
  \bibinfo {pages} {024026} (\bibinfo {year} {2017})}\BibitemShut {NoStop}%
\end{thebibliography}%


\begin{thebibliography}{10}%
\makeatletter
\providecommand \@ifxundefined [1]{%
 \@ifx{#1\undefined}
}%
\providecommand \@ifnum [1]{%
 \ifnum #1\expandafter \@firstoftwo
 \else \expandafter \@secondoftwo
 \fi
}%
\providecommand \@ifx [1]{%
 \ifx #1\expandafter \@firstoftwo
 \else \expandafter \@secondoftwo
 \fi
}%
\providecommand \natexlab [1]{#1}%
\providecommand \enquote  [1]{``#1''}%
\providecommand \bibnamefont  [1]{#1}%
\providecommand \bibfnamefont [1]{#1}%
\providecommand \citenamefont [1]{#1}%
\providecommand \href@noop [0]{\@secondoftwo}%
\providecommand \href [0]{\begingroup \@sanitize@url \@href}%
\providecommand \@href[1]{\@@startlink{#1}\@@href}%
\providecommand \@@href[1]{\endgroup#1\@@endlink}%
\providecommand \@sanitize@url [0]{\catcode `\\12\catcode `\$12\catcode
  `\&12\catcode `\#12\catcode `\^12\catcode `\_12\catcode `\%12\relax}%
\providecommand \@@startlink[1]{}%
\providecommand \@@endlink[0]{}%
\providecommand \url  [0]{\begingroup\@sanitize@url \@url }%
\providecommand \@url [1]{\endgroup\@href {#1}{\urlprefix }}%
\providecommand \urlprefix  [0]{URL }%
\providecommand \Eprint [0]{\href }%
\providecommand \doibase [0]{https://doi.org/}%
\providecommand \selectlanguage [0]{\@gobble}%
\providecommand \bibinfo  [0]{\@secondoftwo}%
\providecommand \bibfield  [0]{\@secondoftwo}%
\providecommand \translation [1]{[#1]}%
\providecommand \BibitemOpen [0]{}%
\providecommand \bibitemStop [0]{}%
\providecommand \bibitemNoStop [0]{.\EOS\space}%
\providecommand \EOS [0]{\spacefactor3000\relax}%
\providecommand \BibitemShut  [1]{\csname bibitem#1\endcsname}%
\let\auto@bib@innerbib\@empty
\bibitem [{E6_()}]{E6_website}%
  \BibitemOpen
  \href@noop {} {\bibinfo {title} {{E}lement {S}ix {T}echnologies {L}td.
  ({UK}), {S}{C} {P}late {C}{V}{D} 3.0x3.0mm, 0.25mm thick, {P}{L}}},\ \bibinfo
  {howpublished}
  {\url{https://e6cvd.com/eur/material/single-crystalline/sc-plate-cvd-3-0x3-0x0-25mm-100-pl.html}}\BibitemShut
  {NoStop}%
\bibitem [{\citenamefont {Mouradian}\ \emph {et~al.}(2017)\citenamefont
  {Mouradian}, \citenamefont {Wan}, \citenamefont {Schröder},\ and\
  \citenamefont {Englund}}]{mouradian_rectangular_2017}%
  \BibitemOpen
  \bibfield  {author} {\bibinfo {author} {\bibfnamefont {S.}~\bibnamefont
  {Mouradian}}, \bibinfo {author} {\bibfnamefont {N.~H.}\ \bibnamefont {Wan}},
  \bibinfo {author} {\bibfnamefont {T.}~\bibnamefont {Schröder}},\ and\
  \bibinfo {author} {\bibfnamefont {D.}~\bibnamefont {Englund}},\ }\href
  {https://doi.org/10.1063/1.4992118} {\bibfield  {journal} {\bibinfo
  {journal} {Applied Physics Letters}\ }\textbf {\bibinfo {volume} {111}},\
  \bibinfo {pages} {021103} (\bibinfo {year} {2017})}\BibitemShut {NoStop}%
\bibitem [{\citenamefont {Pregnolato}\ \emph {et~al.}(2024)\citenamefont
  {Pregnolato}, \citenamefont {Stucki}, \citenamefont {Bopp}, \citenamefont
  {V.~D.~Hoeven}, \citenamefont {Gokhale}, \citenamefont {Krüger},\ and\
  \citenamefont {Schröder}}]{pregnolato_fabrication_2024}%
  \BibitemOpen
  \bibfield  {author} {\bibinfo {author} {\bibfnamefont {T.}~\bibnamefont
  {Pregnolato}}, \bibinfo {author} {\bibfnamefont {M.~E.}\ \bibnamefont
  {Stucki}}, \bibinfo {author} {\bibfnamefont {J.~M.}\ \bibnamefont {Bopp}},
  \bibinfo {author} {\bibfnamefont {M.~H.}\ \bibnamefont {V.~D.~Hoeven}},
  \bibinfo {author} {\bibfnamefont {A.}~\bibnamefont {Gokhale}}, \bibinfo
  {author} {\bibfnamefont {O.}~\bibnamefont {Krüger}},\ and\ \bibinfo {author}
  {\bibfnamefont {T.}~\bibnamefont {Schröder}},\ }\href
  {https://doi.org/10.1063/5.0186509} {\bibfield  {journal} {\bibinfo
  {journal} {APL Photonics}\ }\textbf {\bibinfo {volume} {9}},\ \bibinfo
  {pages} {036105} (\bibinfo {year} {2024})}\BibitemShut {NoStop}%
\bibitem [{\citenamefont {Bopp}\ \emph {et~al.}(2024)\citenamefont {Bopp},
  \citenamefont {Plock}, \citenamefont {Turan}, \citenamefont {Pieplow},
  \citenamefont {Burger},\ and\ \citenamefont {Schröder}}]{bopp_sawfish_2024}%
  \BibitemOpen
  \bibfield  {author} {\bibinfo {author} {\bibfnamefont {J.~M.}\ \bibnamefont
  {Bopp}}, \bibinfo {author} {\bibfnamefont {M.}~\bibnamefont {Plock}},
  \bibinfo {author} {\bibfnamefont {T.}~\bibnamefont {Turan}}, \bibinfo
  {author} {\bibfnamefont {G.}~\bibnamefont {Pieplow}}, \bibinfo {author}
  {\bibfnamefont {S.}~\bibnamefont {Burger}},\ and\ \bibinfo {author}
  {\bibfnamefont {T.}~\bibnamefont {Schröder}},\ }\href
  {https://doi.org/10.1002/adom.202301286} {\bibfield  {journal} {\bibinfo
  {journal} {Advanced Optical Materials}\ }\textbf {\bibinfo {volume} {12}},\
  \bibinfo {pages} {2301286} (\bibinfo {year} {2024})}\BibitemShut {NoStop}%
\bibitem [{\citenamefont {Binder}\ \emph {et~al.}(2017)\citenamefont {Binder},
  \citenamefont {Stark}, \citenamefont {Tomek}, \citenamefont {Scheuer},
  \citenamefont {Frank}, \citenamefont {Jahnke}, \citenamefont {Müller},
  \citenamefont {Schmitt}, \citenamefont {Metsch}, \citenamefont {Unden},
  \citenamefont {Gehring}, \citenamefont {Huck}, \citenamefont {Andersen},
  \citenamefont {Rogers},\ and\ \citenamefont {Jelezko}}]{binder_qudi_2017}%
  \BibitemOpen
  \bibfield  {author} {\bibinfo {author} {\bibfnamefont {J.~M.}\ \bibnamefont
  {Binder}}, \bibinfo {author} {\bibfnamefont {A.}~\bibnamefont {Stark}},
  \bibinfo {author} {\bibfnamefont {N.}~\bibnamefont {Tomek}}, \bibinfo
  {author} {\bibfnamefont {J.}~\bibnamefont {Scheuer}}, \bibinfo {author}
  {\bibfnamefont {F.}~\bibnamefont {Frank}}, \bibinfo {author} {\bibfnamefont
  {K.~D.}\ \bibnamefont {Jahnke}}, \bibinfo {author} {\bibfnamefont
  {C.}~\bibnamefont {Müller}}, \bibinfo {author} {\bibfnamefont
  {S.}~\bibnamefont {Schmitt}}, \bibinfo {author} {\bibfnamefont {M.~H.}\
  \bibnamefont {Metsch}}, \bibinfo {author} {\bibfnamefont {T.}~\bibnamefont
  {Unden}}, \bibinfo {author} {\bibfnamefont {T.}~\bibnamefont {Gehring}},
  \bibinfo {author} {\bibfnamefont {A.}~\bibnamefont {Huck}}, \bibinfo {author}
  {\bibfnamefont {U.~L.}\ \bibnamefont {Andersen}}, \bibinfo {author}
  {\bibfnamefont {L.~J.}\ \bibnamefont {Rogers}},\ and\ \bibinfo {author}
  {\bibfnamefont {F.}~\bibnamefont {Jelezko}},\ }\href
  {https://doi.org/10.1016/j.softx.2017.02.001} {\bibfield  {journal} {\bibinfo
   {journal} {SoftwareX}\ }\textbf {\bibinfo {volume} {6}},\ \bibinfo {pages}
  {85} (\bibinfo {year} {2017})}\BibitemShut {NoStop}%
\bibitem [{\citenamefont {Bopp}\ and\ \citenamefont
  {Schröder}(2024)}]{bopp_dynexphighly_2024}%
  \BibitemOpen
  \bibfield  {author} {\bibinfo {author} {\bibfnamefont {J.~M.}\ \bibnamefont
  {Bopp}}\ and\ \bibinfo {author} {\bibfnamefont {T.}~\bibnamefont
  {Schröder}},\ }\href {https://doi.org/10.1016/j.softx.2024.101964}
  {\bibfield  {journal} {\bibinfo  {journal} {SoftwareX}\ }\textbf {\bibinfo
  {volume} {28}},\ \bibinfo {pages} {101964} (\bibinfo {year}
  {2024})}\BibitemShut {NoStop}%
\bibitem [{\citenamefont {Edmonds}\ \emph {et~al.}(2012)\citenamefont
  {Edmonds}, \citenamefont {D'Haenens-Johansson}, \citenamefont {Cruddace},
  \citenamefont {Newton}, \citenamefont {Fu}, \citenamefont {Santori},
  \citenamefont {Beausoleil}, \citenamefont {Twitchen},\ and\ \citenamefont
  {Markham}}]{Edmonds_production_2012}%
  \BibitemOpen
  \bibfield  {author} {\bibinfo {author} {\bibfnamefont {A.~M.}\ \bibnamefont
  {Edmonds}}, \bibinfo {author} {\bibfnamefont {U.~F.~S.}\ \bibnamefont
  {D'Haenens-Johansson}}, \bibinfo {author} {\bibfnamefont {R.~J.}\
  \bibnamefont {Cruddace}}, \bibinfo {author} {\bibfnamefont {M.~E.}\
  \bibnamefont {Newton}}, \bibinfo {author} {\bibfnamefont {K.-M.~C.}\
  \bibnamefont {Fu}}, \bibinfo {author} {\bibfnamefont {C.}~\bibnamefont
  {Santori}}, \bibinfo {author} {\bibfnamefont {R.~G.}\ \bibnamefont
  {Beausoleil}}, \bibinfo {author} {\bibfnamefont {D.~J.}\ \bibnamefont
  {Twitchen}},\ and\ \bibinfo {author} {\bibfnamefont {M.~L.}\ \bibnamefont
  {Markham}},\ }\href {https://doi.org/10.1103/PhysRevB.86.035201} {\bibfield
  {journal} {\bibinfo  {journal} {Phys. Rev. B}\ }\textbf {\bibinfo {volume}
  {86}},\ \bibinfo {pages} {035201} (\bibinfo {year} {2012})}\BibitemShut
  {NoStop}%
\bibitem [{\citenamefont {Orphal-Kobin}\ \emph {et~al.}(2025)\citenamefont
  {Orphal-Kobin}, \citenamefont {Pieplow}, \citenamefont {Gokhale},
  \citenamefont {Unterguggenberger},\ and\ \citenamefont
  {Schröder}}]{orphal-kobin_retrieving_2025}%
  \BibitemOpen
  \bibfield  {author} {\bibinfo {author} {\bibfnamefont {L.}~\bibnamefont
  {Orphal-Kobin}}, \bibinfo {author} {\bibfnamefont {G.}~\bibnamefont
  {Pieplow}}, \bibinfo {author} {\bibfnamefont {A.}~\bibnamefont {Gokhale}},
  \bibinfo {author} {\bibfnamefont {K.}~\bibnamefont {Unterguggenberger}},\
  and\ \bibinfo {author} {\bibfnamefont {T.}~\bibnamefont {Schröder}},\ }\href
  {https://doi.org/10.48550/ARXIV.2501.07951} {\bibinfo {title} {Retrieving
  {Lost} {Atomic} {Information}: {Monte} {Carlo}-based {Parameter}
  {Reconstruction} of an {Optical} {Quantum} {System}}} (\bibinfo {year}
  {2025}),\ \Eprint {https://arxiv.org/abs/2501.07951} {arXiv:2501.07951
  [quant-ph]} \BibitemShut {NoStop}%
\bibitem [{jcm()}]{jcmwave}%
  \BibitemOpen
  \href@noop {} {\bibinfo {title} {{J}{C}{M}suite}},\ \bibinfo {howpublished}
  {\url{https://jcmwave.com/jcmsuite}},\ \bibinfo {note} {{J}CMwave GmbH,
  Bolivarallee 22, 14050 Berlin, Germany}\BibitemShut {NoStop}%
\bibitem [{\citenamefont {Maze}\ \emph {et~al.}(2011)\citenamefont {Maze},
  \citenamefont {Gali}, \citenamefont {Togan}, \citenamefont {Chu},
  \citenamefont {Trifonov}, \citenamefont {Kaxiras},\ and\ \citenamefont
  {Lukin}}]{Maze_properties_2011}%
  \BibitemOpen
  \bibfield  {author} {\bibinfo {author} {\bibfnamefont {J.~R.}\ \bibnamefont
  {Maze}}, \bibinfo {author} {\bibfnamefont {A.}~\bibnamefont {Gali}}, \bibinfo
  {author} {\bibfnamefont {E.}~\bibnamefont {Togan}}, \bibinfo {author}
  {\bibfnamefont {Y.}~\bibnamefont {Chu}}, \bibinfo {author} {\bibfnamefont
  {A.}~\bibnamefont {Trifonov}}, \bibinfo {author} {\bibfnamefont
  {E.}~\bibnamefont {Kaxiras}},\ and\ \bibinfo {author} {\bibfnamefont {M.~D.}\
  \bibnamefont {Lukin}},\ }\href
  {https://doi.org/10.1088/1367-2630/13/2/025025} {\bibfield  {journal}
  {\bibinfo  {journal} {New Journal of Physics}\ }\textbf {\bibinfo {volume}
  {13}},\ \bibinfo {pages} {025025} (\bibinfo {year} {2011})}\BibitemShut
  {NoStop}%
\end{thebibliography}%
